\documentclass[aip,reprint,twocolumn,nofootinbib]{revtex4}

\usepackage[]{graphicx}
\usepackage{amsmath,amssymb}
\usepackage{bm}
\usepackage{color}
\usepackage[normalem]{ulem}
\usepackage[usenames,dvipsnames]{xcolor}
\usepackage{eufrak}
\usepackage{bbm}

\DeclareMathOperator{\sgn}{sgn}

\begin{document}

\title{Theory of light-activated catalytic Janus particles}
\date{\today}

\author{W. E. Uspal}
\affiliation{Department of Mechanical Engineering, University of Hawai'i at Manoa, 2540 Dole Street,
Holmes 302, Honolulu, HI 96822}
\affiliation{Max-Planck-Institut f\"{u}r Intelligente Systeme, 
Heisenbergstr. 3, D-70569, Stuttgart, Germany}
\affiliation{IV. Institut f\"ur Theoretische Physik, 
Universit\"{a}t Stuttgart, Pfaffenwaldring 57, D-70569 Stuttgart, Germany}
\email{uspal@hawaii.edu}

\begin{abstract}
We study the dynamics of active Janus particles that self-propel in solution by light-activated catalytic decomposition of chemical ``fuel.''  We develop an analytical model of a photo-active self-phoretic particle that accounts for ``self-shadowing'' of the light by the opaque catalytic face of the particle. We find that self-shadowing can drive ``phototaxis" (rotation of the catalytic cap towards the light source) or ``anti-phototaxis," depending on the properties of the particle. Incorporating the effect of thermal noise, we show that the distribution of particle orientations is captured by a Boltzmann distribution with a nonequilibrium effective potential. Furthermore, the mean vertical velocity of phototactic (anti-phototactic) particles exhibits a superlinear (sublinear) dependence on intensity. Overall, our findings show that photo-active particles exhibit a rich ``tactic''' response to light, which could be harnessed to program complex three-dimensional trajectories. \end{abstract}

\maketitle

\section{Introduction}

Chemically active colloids are capable of transducing chemical free energy, available as molecular ``fuel'' diffusing in the surrounding solution, into mechanical motion of the solution and (usually) of the colloid itself.\cite{moran17} These particles have far-reaching potential applications that originate in their intrinsically non-equilibrium character. For example, they could serve as ``engines''  for  micro- and nano-robots that operate autonomously in micro-confined spaces, or as building blocks of dissipative materials that mimic emergent features of biological systems. Consequently, significant effort has been invested in the development of a large library of active colloid designs.\cite{sanchez15,aubret17} Colloidal self-propulsion has been demonstrated for a  variety of particle geometries (e.g.,  bent,\cite{tenhagen14,oneeljudy19} tree-like,\cite{dai16} and hollow\cite{ma17}), reaction schemes (e.g., with hydrogen peroxide or quinones\cite{dai16} as the ``fuel''), and material compositions. 

An emerging focus area in active colloids research is the design and fabrication of active colloids that are responsive to specific ambient fields.\cite{stark16, zottl16} As one motivation for this work, these colloids could be guided with fields that are applied and controlled externally, opening new possibilities in targeted micro-cargo transport \cite{sundararajan08,palacci13b,chen17} and assembly of micromachines and dissipative materials.\cite{maggi15,aubret18} To this end, magnetic materials have been incorporated within catalytic Janus particles for external control over the orientations and trajectories of individual particles.\cite{sundararajan08,baraban12, nourhani17} Secondly, field-responsive colloids could potentially harness the directional information inherent in ambient fields in order to navigate long distances through complex environments. In analogy with biological systems, this directed spatial migration can be considered a form of ``taxis''.   Recent studies have investigated the  rheotactic response of chemically active particles to hydrodynamic flow,\cite{palacci15,uspal15b,ren17,katuri18} gravitactic response to the earth's gravitational field,\cite{campbell13,tenhagen14,campbell17} ``viscotactic'' response to viscosity gradients,\cite{liebchen18} chemotactic response to gradients in the concentration of chemical ``fuel'',\cite{hong07, baraban13, saha14,byun17,geiseler17,jin17,chen18,tatulea18,popescu18NL} and ``thigmotactic'' response to gradients in the material composition of bounding surfaces\cite{uspal16,popescu17b,Holterhoff18}. These various forms of taxis can be understood on the basis of the microscopic physics of how the ambient field couples to the activity and motion of the particle.\cite{saha14,bickel14,tatulea18,popescu18NL}

In this context, particular attention has been devoted to the development of photo-active colloids. \cite{hong10, palacci13, palacci13b,dai16,altemose17,chen17,wang18,singh18,aubret18,oneeljudy19} Typically, the catalytic region of a photo-active particle is made of a semiconducting material, and catalyzes the decomposition of molecular ``fuel'' only when exposed to light of wavelength corresponding to the material's electronic bandgap. Therefore, the trajectory of a photo-active particle can be controlled through the intensity and direction of an external illumination source.\cite{dai16,lozano16,chen17,aubret18} Additionally, these particles could potentially harness variations in ambient light for phototactic navigation, mimicking certain micro-organisms.\cite{garcia13, giometto15} In some of these cases, the materials comprising the particle are  transparent or near-transparent to the incident light. However, if the materials have limited penetration depth (e.g., for semiconducting materials), and if the particle size is comparable to or larger than the wavelength of incident light, the particle can be regarded as opaque, and rich new physics will arise from a ``self-shadowing'' effect. For instance,  an opaque spherical particle with a uniform catalytic surface will behave as an effective ``Janus'' particle when exposed to light, with its axis of symmetry always aligned with the direction of illumination.\cite{cohen14,li16,chen17} The hemispherical half of the particle closest to the light will catalyze the reaction, while the other half will be in shadow and therefore inactive. If the particle is intrinsically a Janus particle, in the sense that its surface comprises distinct regions of catalytic and inert material, then the distribution of surface activity will have a complicated dependence on the orientation,  with respect to the direction of incident light, of the particle's axis of symmetry. Moreover, if the various materials comprising the particle surface have different potentials of interaction with the various molecular species  involved in the reaction, the particle will rotate in response to light, i.e., exhibit a phototactic response.\cite{singh18}

In this paper, we develop a theory describing phototaxis of a self-shadowing photo-active Janus particle. We model the hydrodynamic and chemical fields created by the particle as continua, and treat occlusion of the incident light by the particle surface through simple geometric optics. Analytically and numerically, we find that a half-covered particle will  rotate its catalytic cap towards (phototaxis) or away from (anti-phototaxis) a light source, depending on the surface chemistry of the particle. We also obtain the translational velocity of the particle as a function of orientation. Numerically investigating the effect of the extent of coverage by catalyst, we find that it can significantly change the form, as a function of particle orientation, of the light-induced angular velocity. In particular, for low coverage particles, this function exhibits a marked departure from a sinusoidal form, owing to the broad range of orientations in which the catalytic cap is completely shadowed.

 In addition to these deterministic effects, we  consider the role of thermal fluctuations. Analytically and numerically, we find that the probability distribution of particle orientations is captured by a Boltzmann distribution with a non-equilibrium potential. We show that, for phototactic (anti-photactic) particles, the mean vertical velocity exhibits a superlinear (sublinear) dependence on particle activity. We also consider the dynamical phase behavior, i.e., whether the particle, on average, sediments or swims vertically. For anti-phototactic particles, the interplay of bottom-heaviness and swimming activity can lead to re-entrant behavior, providing a clear signature of anti-phototaxis for experimental studies. Overall, our findings illustrate the rich physics that can arise from the microscopic coupling of an ambient optical field to chemical activity of a colloid.

\section{Deterministic theory}
\subsection{Model for particle activity}
\begin{figure*}[thb]
\includegraphics[scale=0.7]{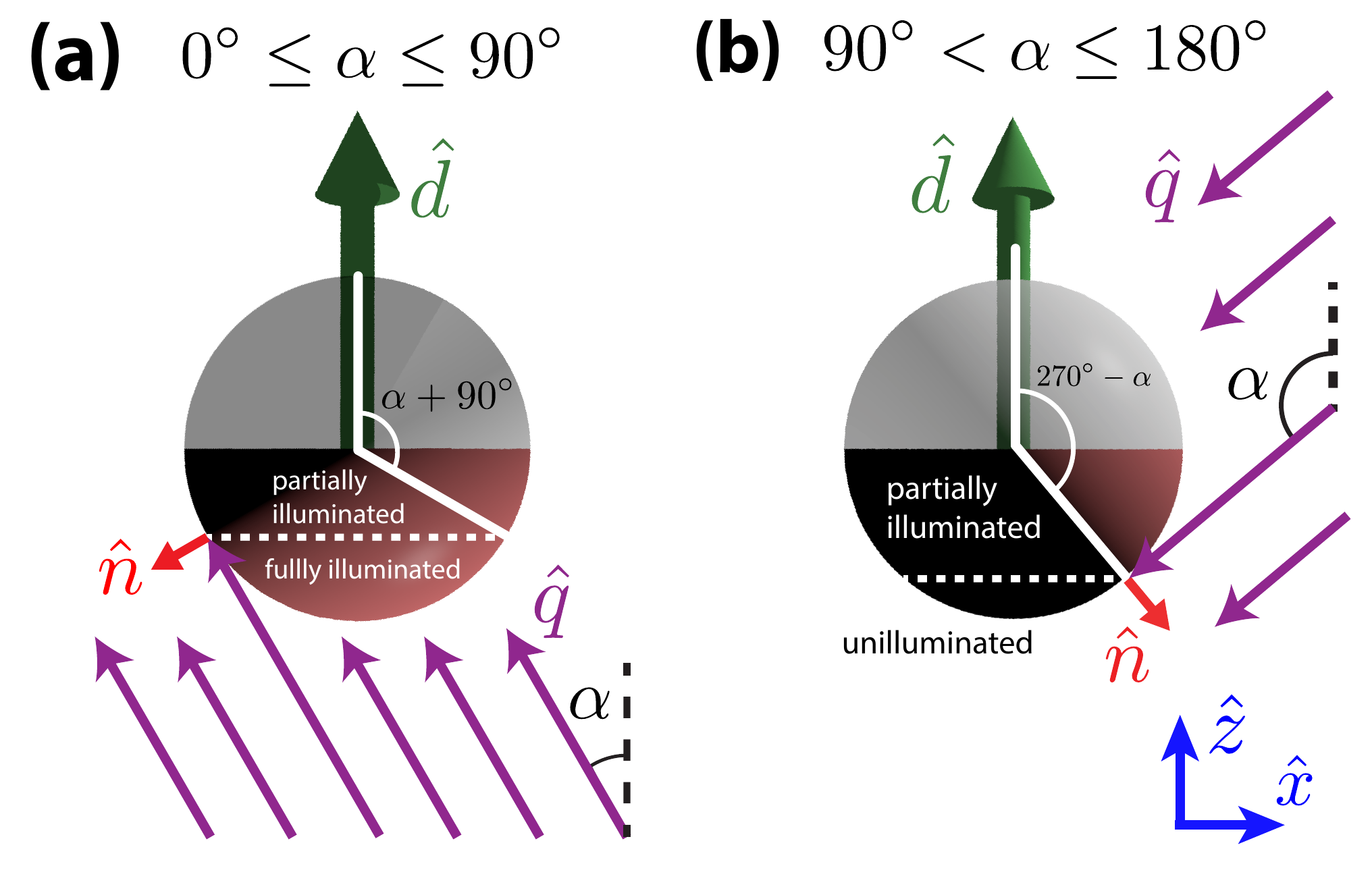}
\caption{\label{fig:combined_schematic} Schematic illustrating a light-activated catalytic Janus particle in uniform plane illumination. Here, the particle axis $\mathbf{\hat{d}}$ makes an angle $\alpha$ with respect to the direction $\mathbf{\hat{q}}$ of the light. We assume a coordinate frame that is co-moving with the particle, such that $\hat{z}$ is parallel to $\mathbf{\hat{d}}$. The surface of the particle is described with spherical coordinates $\theta$ and $\phi$. (a) \textcolor{black}{The case of $0 \leq \alpha \leq 90^{\circ}$.} Part of the catalytic cap  \textcolor{black}{($90^{\circ} + \alpha \leq \theta \leq 180^{\circ}$)} is completely illuminated by light. There is also a partially illuminated region, defined by \textcolor{black}{$90^{\circ} \leq \theta \leq 90^{\circ} + \alpha$}. In this region, only the range of angles $-\phi_0(\theta, \alpha) < \phi < \phi_0(\theta, \alpha)$ receive light,  where $\phi_0(\theta, \alpha) = \arccos(\cot \theta \cot \alpha)$. (b) \textcolor{black}{The case of $90^{\circ} < \alpha \leq 180^{\circ}$.} A region of the catalytic cap \textcolor{black}{($270^{\circ} - \alpha \leq \theta < 180^{\circ}$)} is completely in shadow. Another region \textcolor{black}{($90^{\circ} \leq \theta < 270^{\circ} - \alpha$)} is illuminated between the angles $-\phi_0(\theta, \alpha) < \phi < \phi_0(\theta, \alpha)$. }
\end{figure*}

We consider a spherical, light-activated catalytic Janus particle of radius $R$ in unbounded solution.  The particle is half-covered by catalyst, and the orientation of the particle is described by the vector $\mathbf{\hat{d}}$, which lies along the axis of the symmetry of the particle, and is defined to point from the catalytic pole to the inert pole. Light with uniform intensity $I$ is shining on the particle with direction of propagation $\mathbf{\hat{q}}$, where the angle between $\mathbf{\hat{d}}$ and $\mathbf{\hat{q}}$ is $\alpha$, as shown in \textcolor{black}{Fig. \ref{fig:combined_schematic}}. Where the catalyst is illuminated, the local flux of product molecules (``solute'') is proportional to the local flux of incident light. The solute diffuses in the surrounding solution with a diffusion constant $D$.  The solute diffuses very fast, relative to the motion of the particle, i.e., the P\'{e}clet number $Pe \equiv U_0 R/D$ is small, where $U_0$ is a characteristic self-propulsion velocity that will be defined later. Accordingly, the solute number density field $c(\mathbf{x})$ is governed by the Laplace equation, $D \nabla^{2} c(\mathbf{x}) = 0$, where $\mathbf{x}$ is a point in the liquid solution. 
The boundary condition on the solute number density field is
\textcolor{black}{
\begin{equation}
\label{eq:bc}
-D \left[ \bm{\hat{n}} \cdot \nabla c \right] = \kappa \, (-\bm{\hat{n}} \cdot \mathbf{\hat{q}}) \, \Theta(-\bm{\hat{n}} \cdot \mathbf{\hat{q}}) \, \Theta(-\bm{\hat{n}} \cdot \mathbf{\hat{d}})
\end{equation}}
over the surface of the particle. Here, $\Theta(x)$ is the Heaviside step function, and the local surface normal $\mathbf{\hat{n}}$ is defined to point from the particle surface into the liquid solution. In the boundary condition, the first step function represents the condition that only illuminated areas produce solute, while the second step function represents the condition that this only occurs on the catalytic cap.  The rate of solute production per unit area  $\kappa$ is some function of the incident light intensity $I$,  $\kappa = \kappa(I)$, with $\kappa(0) = 0$.  Additionally, the solute number density field decays to a constant value $c^{\infty}$ far away from the particle.

Without loss of generality, we specify that the orientation vector $\mathbf{\hat{d}}$ lies in the $\hat{z}$ direction, and $\mathbf{\hat{q}}$ is the in $xz$ plane with $-1 \leq {q}_{x} \leq 0$ and $-1 \leq q_{z} \leq 1$. The $\mathbf{\hat{q}}$ vector has an angle $\alpha$ with respect to the $\hat{z}$ direction. We introduce a spherical coordinate system with its origin at the particle center.  We expand the solute number density field in spherical harmonics:
\begin{equation}
\begin{split}
c(r, \theta, \phi) = c^{\infty} +   \frac{\kappa R}{D} & \sum_{l=0}^{\infty}  \sum_{m = 0}^{l}  \frac{1}{l+1} \left(\frac{R}{r}\right)^{l + 1} P_{l}^{m}(\cos(\theta)) \\ &  (A_{lm} \cos(m \phi) + B_{lm} \sin(m \phi)).  
\label{eq:SH_expansion}
\end{split}
\end{equation}
The coefficients $A_{lm}$ and $B_{lm}$ are dimensionless. Due to the mirror symmetry across the $xz$ plane, the coefficients $B_{lm}$ all vanish. Additionally, if $\alpha = 0$, then the geometry is axisymmetric; all the coefficients with $m \neq 0$ vanish, and the problem is identical to the ``catalyst-thickness-dependent'' case previously analyzed in Refs. \citenum{popescu17} and \citenum{popescu18}. Of course, if \textcolor{black}{$\alpha = 180^{\circ}$}, we obtain the trivial case of completely shadowed catalyst and no particle motion. For the general case of \textcolor{black}{$0^{\circ} < \alpha < 180^{\circ}$}, we need to find the set $A_{lm}$.  We take the derivative $-D \left( \frac{\partial c}{\partial r} \right)|_{r = R}$ of Eq. (\ref{eq:SH_expansion}) \textcolor{black}{and apply the boundary condition given in Eq. (\ref{eq:bc}):}
\begin{multline}
\kappa \, (-\bm{\hat{n}} \cdot \mathbf{\hat{q}}) \, \Theta(-\bm{\hat{n}} \cdot \mathbf{\hat{q}}) \, \Theta(-\bm{\hat{n}} \cdot \mathbf{\hat{d}}) = \\  \kappa \sum_{l = 0}^{\infty} \sum_{m = 0}^{\infty} A_{lm} P_{l}^{m} (\cos \theta) \cos(m \phi).
\end{multline}
We use the orthogonality of spherical harmonics to match the boundary conditions:
\begin{multline}
\int_{-\pi}^{\pi} d \phi \int_{0}^{\pi} d \theta \sin \theta \, (-\bm{\hat{n}} \cdot \mathbf{\hat{q}}) \, \Theta(-\bm{\hat{n}} \cdot \mathbf{\hat{q}}) \, \Theta(-\bm{\hat{n}} \cdot \mathbf{\hat{d}}) \\  P_{l'}^{m'} (\cos \theta) \cos(m' \phi)  \\ =
\sum_{l = 0}^{\infty} \sum_{m = 0}^{\infty} A_{lm} \int_{-\pi}^{\pi} d \phi \int_{0}^{\pi} d \theta \sin \theta   P_{l}^{m} (\cos \theta) P_{l'}^{m'} (\cos \theta) \\  \cos(m \phi) \cos(m' \phi).
\end{multline}
\begin{multline}
\int_{-\pi}^{\pi} d \phi \int_{0}^{\pi} d \theta \sin \theta \, (-\bm{\hat{n}} \cdot \mathbf{\hat{q}}) \, \Theta(-\bm{\hat{n}} \cdot \mathbf{\hat{q}}) \, \Theta(-\bm{\hat{n}} \cdot \mathbf{\hat{d}}) \\ P_{l'}^{m'} (\cos \theta) \cos(m' \phi)  =  A_{l'm'}  
\frac{2 \pi (1 + \delta_{m'0}) (l' + m')!}{(2l' + 1)(l' - m')!}.
\end{multline}
Dropping the primes:
\begin{multline}
A_{lm} = \frac{(2l + 1)(l - m)!}{2 \pi (1 + \delta_{m0}) (l + m)!} \\ \int_{-\pi}^{\pi} d \phi \int_{0}^{\pi} d \theta \sin \theta \, (-\bm{\hat{n}} \cdot \mathbf{\hat{q}}) \, \Theta(-\bm{\hat{n}} \cdot \mathbf{\hat{q}}) \, \Theta(-\bm{\hat{n}} \cdot \mathbf{\hat{d}}) \\ P_{l}^{m} (\cos \theta) \cos(m \phi).
\end{multline}
We also have:
\begin{equation}
\mathbf{\hat{n}} \cdot \mathbf{\hat{q}} = \cos(\alpha) \cos(\theta) - \sin(\alpha) \sin(\theta) \cos(\phi).
\end{equation}

Before performing the integration, we distinguish two scenarios for the self-shadowing of a half-covered particle. As the first scenario, we consider angles of incidence in the range \textcolor{black}{$0^{\circ} \leq \alpha < 90^{\circ}$}. In this scenario, the catalytic cap has two distinct subregions \textcolor{black}{(Fig. \ref{fig:combined_schematic}(a))}. For the region defined by the polar angles \textcolor{black}{$90^{\circ} + \alpha < \theta \leq 180^{\circ}$}, the catalyst receives incident light over the entire azimuthal range \textcolor{black}{$-180^{\circ} \leq \phi \leq 180^{\circ}$}. We call this the ``fully illuminated'' region. For the region defined by \textcolor{black}{$90^{\circ} < \theta \leq 90^{\circ} + \alpha$}, the catalyst receives incident light over a range of azimuthal angles $[-\phi_0, \phi_0]$. We call this range of polar angles the ``partially illuminated'' region. For a polar angle $\theta$ in the partially illuminated region, the angle $\phi_0$ is defined as the azimuthal angle where the normal flux of incident light is zero:

\begin{equation}
 \cos(\alpha) \cos(\theta) - \sin(\alpha) \sin(\theta) \cos(\phi_0) \equiv 0,
\end{equation}
so that
\begin{equation}
\phi_{0} = \arccos(\cot \theta \cot \alpha).
\end{equation}

The second scenario occurs when the incidence angle $\alpha$ is in the range $90^{\circ} < \alpha \leq 180^{\circ}$ \textcolor{black}{(Fig. \ref{fig:combined_schematic}(b))}. In this case, a region of the cap defined by $270^{\circ} - \alpha < \theta \leq 180^{\circ}$ is completely in shadow, i.e., the normal flux of incident light is zero for all azimuthal angles $\phi$. The region $90^{\circ} \leq \theta \leq 270^{\circ} - \alpha$ is partially illuminated, with light falling on the range of azimuthal angles $[-\phi_0, \phi_0]$, with $\phi$ defined as before.

In the following subsections, our aim is to provide analytical expressions for the coefficients $A_{lm}$. We focus on $m = 0$ and $m = 1$, as these will be needed to calculate the translational and angular velocity of a swimming particle. We distinguish $A_{lm} \equiv A_{lm}^{(lt90)}$ for $0^{\circ} \leq  \alpha \leq 90^{\circ}$ and $A_{lm} \equiv A_{lm}^{(gt90)}$ for $90^{\circ} < \alpha \leq 180^{\circ}$.

\subsubsection{First scenario: \textcolor{black}{$0^{\circ} \leq \alpha \leq 90^{\circ}$}}

We will consider the contributions of the partially illuminated and fully illuminated regions separately, and define:
\begin{equation}
A_{lm}^{(lt90)} \equiv -\frac{(2l + 1)(l - m)!}{2 \pi (1 + \delta_{m0}) (l + m)!}  \left[A_{lm}^{(lt90,pi)} + A_{lm}^{(lt90,fi)} \right].
\end{equation}
The contribution of the partially illuminated region is
\begin{multline}
A_{lm}^{(lt90,pi)} = 
\int_{-\phi_0}^{\phi_0} d \phi \int_{\pi/2}^{\pi/2 + \alpha} d \theta \sin \theta \,  [\cos(\alpha) \cos(\theta) - \\ \sin(\alpha) \sin(\theta) \cos(\phi)] P_{l}^{m} (\cos \theta) \cos(m \phi).
\end{multline}
For the fully illuminated region,
\begin{multline}
A_{lm}^{(lt90,fi)} = 
\int_{0}^{2 \pi} d \phi \int_{\pi/2 + \alpha}^{\pi} d \theta \sin \theta \,  [\cos(\alpha) \cos(\theta) - \\  \sin(\alpha) \sin(\theta) \cos(\phi)] P_{l}^{m} (\cos \theta) \cos(m \phi).
\end{multline}
Evaluating the two integrals, we obtain the coefficients $A_{lm}^{(lt90)}$ given in Appendix A.

\subsubsection{Second scenario: \textcolor{black}{$90^{\circ} < \alpha \leq 180^{\circ}$}}

In this scenario, shown schematically by Fig. \ref{fig:combined_schematic}(b), only one integral contributes to $A_{lm}$:
\begin{multline}
A_{lm}^{(gt90)} = -\frac{ (2l + 1)(l - m)!}{2 \pi  \textcolor{black}{(1 + \delta_{m0})} (l + m)!}  \, \textcolor{black}{A_{lm}^{(gt90,pi)}},
\end{multline}
\textcolor{black}{with
\begin{multline}
A_{lm}^{(gt90,pi)} = \int_{-\phi_0}^{\phi_0} d \phi \int_{\pi/2}^{3 \pi/2 - \alpha } d \theta \sin \theta \, [\cos(\alpha) \cos(\theta) - \\  \sin(\alpha) \sin(\theta) \cos(\phi)] P_{l}^{m} (\cos \theta) \cos(m \phi).
\end{multline}}
As with the first scenario, we have evaluated the coefficients $A_{lm}^{(gt90)}$ for $l \in \{1, ..., 5\}$ and $m \in \{0, 1\}$. \textcolor{black}{Surprisingly, the expressions for $A_{lm}^{(lt90)}$ and for $A_{lm}^{(gt90)}$  evaluate to the same numerical values over the domain $90{^\circ} < \alpha \leq 180^{\circ}$, even though $A_{lm}^{(lt90)}$ was initially obtained for the range $0{^\circ} \leq \alpha \leq 90^{\circ}$.}\footnote{\textcolor{black}{They do not necessarily evaluate to the same values in the range $0{^\circ} \leq \alpha \leq 90^{\circ}$.}} Therefore, we omit the expressions for $A_{lm}^{(gt90)}$ for the sake of brevity, \textcolor{black}{and hereafter take $A_{lm} = A_{lm}^{(lt90)}$ over the whole range of $\alpha$.}

%
%
%
%
%
%
%
%
%
%
%

\subsection{Particle velocity}
 In order to model self-propulsion through liquid driven by chemical gradients, we use the classical theory of diffusiophoresis.\cite{moran17,anderson89,golestanian07}  We take the suspending fluid to be an incompressible Newtonian liquid with mass density $\rho$ and dynamic viscosity \textcolor{black}{$\mu$}. We assume small Reynolds number \textcolor{black}{$Re \equiv \rho U_0 R / \mu$}. Therefore, the fluid velocity $\mathbf{u}(\mathbf{x})$  and pressure $P(\mathbf{x})$ of the solution are governed by the Stokes equation
\textcolor{black}{\begin{equation}
-\nabla P + \mu \nabla^2 \mathbf{u} = 0,
\end{equation}}
and the incompressibility condition $\nabla \cdot \mathbf{u} = 0$. The interaction of the solute molecules with the surface of the particle drives an interfacial flow that we model with an effective slip velocity
\begin{equation}
\label{eq:slip_vel}
\mathbf{v}_s(\mathbf{x}_{s}) = -b(\mathbf{x}_{s}) \nabla_{||} c(\mathbf{x})|_{\mathbf{x} = \mathbf{x}_{s}}.
\end{equation}
Here, $\mathbf{x}_{s}$ is a point on the surface of the particle, $\nabla_{||} \equiv \left( \mathbf{I} - \bm{\hat{n}} \bm{\hat{n}} \right) \cdot \nabla$, and $b(\mathbf{x}_{s})$ is a material dependent parameter (the 
so-called ``surface mobility'') that encapsulates the molecular 
details of the interaction between the solute and the particle surface.\cite{anderson89} We note that $b < 0$ represents an effective repulsive interaction, and $b > 0$ represents an effective 
attractive interaction. The fluid velocity $\mathbf{u}(\mathbf{x})$ satisfies the boundary conditions
\begin{equation}
\mathbf{u}|_{\mathbf{x} = \mathbf{x}_{s}} = \mathbf{U} + \bm{\Omega} \times 
(\mathbf{x}_{s} - \mathbf{x}_{p}) + \mathbf{v}_{s}(\mathbf{x}_{s})
\end{equation}
on the surface of the particle, where $\mathbf{x}_{p}$ is the position of the particle, and $\mathbf{U}$ and $\bm{\Omega}$ are the unknown translational and rotational velocities of the \textcolor{black}{particle}, respectively. Additionally, $\mathbf{u}(\mathbf{x})$ vanishes far away from the particle, i.e., $\mathbf{u}(|\mathbf{x} - \mathbf{x}_{p}| \rightarrow \infty) = 0$. Finally, to the close the system of equations for $\mathbf{U}$ and $\bm{\Omega}$, we write a force balance equation:
\begin{equation}
\int \bm{\sigma} \cdot \mathbf{\hat{n}}\;dS = 0,
\end{equation}
and a torque balance equation
\begin{equation}
\int (\mathbf{x}_s - \mathbf{x}_p) \times \bm{\sigma} \cdot \mathbf{\hat{n}}\;dS  = 0.
\end{equation}
Here, \textcolor{black}{$\bm{\sigma} = -P \,\mathbf{I} + \mu [\nabla \mathbf{u} + \nabla^{T} \mathbf{u}]$} is the stress tensor for a Newtonian liquid, and the integrals are taken over the particle surface. We note that in this Section we assume that there are no external forces or torques on the particle. Since the governing equations are linear, the contribution of any external forces or external torques (e.g., from gravity) to the particle \textcolor{black}{velocity can} be calculated separately, using standard methods, and superposed with the swimming velocities $\mathbf{U}$ and $\bm{\Omega}$ to obtain the complete velocity of the particle. (Section \ref{Sec:Fluctuating} will consider gravitational effects.)

In order to obtain $\mathbf{U}$ and $\bm{\Omega}$, we use the Lorentz reciprocal theorem:\cite{stone96}
\begin{equation}
\mathbf{U} = - \frac{1}{4 \pi R^{2}} \int \mathbf{v}_{s} \, dS
\end{equation}
and 
\begin{equation}
\bm{\Omega} = - \frac{3}{8 \pi R^{3}} \int \mathbf{n} \times \mathbf{v}_{s} \, dS,
\end{equation}
where the integrals are performed over the surface of the particle. Recalling that $\mathbf{\hat{d}} = \hat{z}$, the translational velocity can be written
\begin{equation}
\mathbf{U} = U^{p} \, \hat{x} + U^{d} \, \hat{z},
\end{equation}
where $U^{d}$ is the component of the velocity along the particle axis, and $U^{p}$ is the component perpendicular to the particle axis. By symmetry and the definition of the coordinate system, there is no component in the $\hat{y}$ direction. Furthermore, the particle has an angular velocity
\begin{equation}
\bm{\Omega} = \Omega_{y} \, \hat{y},
\end{equation}
with $\Omega_{x} = \Omega_{z} = 0$ \textcolor{black}{and $\dot{\alpha} = \Omega_y$}.

The slip velocity on the surface of the particle is
\textcolor{black}{
\begin{equation}
\mathbf{v}_{s}(\theta, \phi) = v_{s,\theta} \hat{\theta} + v_{s,\phi} \hat{\phi}.
\end{equation}}
\textcolor{black}{Using Eq. (\ref{eq:slip_vel}), we obtain}
\begin{equation}
\mathbf{v}_{s} = - \frac{b(\theta)}{R} \left[ \frac{\partial c}{\partial \theta} \hat{\theta} + \frac{1}{\sin \theta} \frac{\partial c}{\partial \phi} \hat{\phi} \right],
\end{equation}
where the surface mobility $b(\theta)$ can potentially vary over the surface of the particle. We define
\begin{equation}
b(\theta) \equiv b_0 \, g(\theta),
\end{equation}
such that $g(\theta)$ is dimensionless. We obtain
\begin{equation}
v_{s,\theta} = \textcolor{black}{\frac{\kappa \, b_0 \, g(\theta)}{D}} \, \sum_{l = 1}^{\infty} \sum_{m = 0}^{l}  \frac{A_{lm}}{l+1} \, \sin \theta \, \frac{\partial P_{l}^{m}(\cos \theta)  }{\partial (\cos \theta)}  \, \cos(m \phi),
\end{equation}
which can be rearranged as
\begin{multline}
v_{s,\theta} = \frac{\kappa b_0 \, g(\theta)}{D} \, \sum_{l = 1}^{\infty} \sum_{m = 0}^{l}  \frac{A_{lm}}{l+1} \, \frac{1}{2} \left[(l + m)(l - m + 1) \right. \\ \left. P_{l}^{m-1}(\cos \theta) - P_{l}^{m+1}(\cos \theta) \right]  \, \cos(m \phi),
\end{multline}
\textcolor{black}{where we have used a recurrence relation for $\sqrt{1-x^2} \frac{dP_{l}^{m}(x)}{dx}$.}
\textcolor{black}{Concerning the azimuthal component, we obtain}
\begin{equation}
v_{s,\phi} = \frac{\kappa b_0 \, g(\theta)}{D} \, \sum_{l = 1}^{\infty} \sum_{m = 0}^{l}  \frac{m  A_{lm}}{l+1} \, \frac{1}{\sin \theta} \, P_{l}^{m}(\cos \theta) \, \sin(m \phi).
\end{equation}

\subsubsection{Translation along particle axis}
We first consider the component of translational velocity along the particle axis: 
\begin{multline}
U^{d} = -\frac{1}{4 \pi R^{2}} \int \left[ v_{s,\theta} (\hat{z} \cdot \hat{\theta} ) + v_{s,\phi} (\hat{z} \cdot \hat{\phi}) \right] dS \\ =  \frac{1}{4 \pi R^{2}} \int v_{s,\theta} \sin \theta \, dS.
\end{multline}
\begin{multline}
\begin{split}
U^{d} = \frac{\kappa b_0}{4 \pi D} \,  \sum_{l = 1}^{\infty} \sum_{m = 0}^{l} \int d{\theta} \, \sin \theta \, \int d{\phi} \, g(\theta) \, \frac{A_{lm}}{l+1} \, \\  \frac{1}{2} \left[(l + m)(l - m + 1) P_{l}^{m-1}(\cos \theta) \right. \\ \left.
- P_{l}^{m+1}(\cos \theta) \right]  \, \cos(m \phi) \, \sin \theta.
\end{split}
\end{multline}
\textcolor{black}{We perform the integration over the $\phi$ coordinate. The terms with $m > 0$ vanish by symmetry, giving}
\begin{multline}
U^{d} = \frac{\kappa b_0}{2 D}  \,  \sum_{l = 1}^{\infty} \int d{\theta} \, \sin \theta \, g(\theta) \, \frac{A_{l0}}{l+1} \, \\  \frac{1}{2} \left[l (l + 1) P_{l}^{-1}(\cos \theta) - P_{l}^{1}(\cos \theta) \right]  \sin \theta.
\end{multline}
Using $\sin \theta = -P_{1}^{1} (\cos \theta)$ and
\begin{equation}
P_{l}^{-m} = (-1)^{m} \frac{(l-m)!}{(l + m)!} P_{l}^{m},
\end{equation}
we obtain
\begin{multline}
U^{d} = - \frac{\kappa b_0}{2 D}  \,  \sum_{l = 1}^{\infty} \int d{\theta} \, \sin \theta \, g(\theta) \, \frac{A_{l0}}{l+1} \,   \\ \frac{1}{2} \left[- l (l + 1)  \frac{(l-1)!}{(l + 1)!}  - 1 \right] P_{l}^{1}(\cos \theta)  P_{1}^{1}(\cos \theta),
\end{multline}
\begin{equation}
U^{d} =  \frac{\kappa b_0}{2 D}  \,  \sum_{l = 1}^{\infty}  \, \frac{A_{l0}}{l+1} \, \int d{\theta} \, \sin \theta \, g(\theta) \, P_{l}^{1}(\cos \theta) P_{1}^{1}(\cos \theta).
\end{equation}
Defining
 \begin{equation}
{\cal I}_{l,k}^{m,n} \equiv \frac{1}{2} \int d{\theta} \, \sin \theta \, g(\theta) \, P_{l}^{m}(\cos \theta)  P_{k}^{n}(\cos \theta),
\end{equation}
we have
\begin{equation}
\label{eq:ud}
U^{d} =  \frac{\kappa b_0}{D}  \,  \sum_{l = 1}^{\infty}  \, \frac{A_{l0}}{l+1} \, {\cal I}_{1,l}^{1,1}.
\end{equation}

\begin{figure}[htb]
\includegraphics[width=\columnwidth]{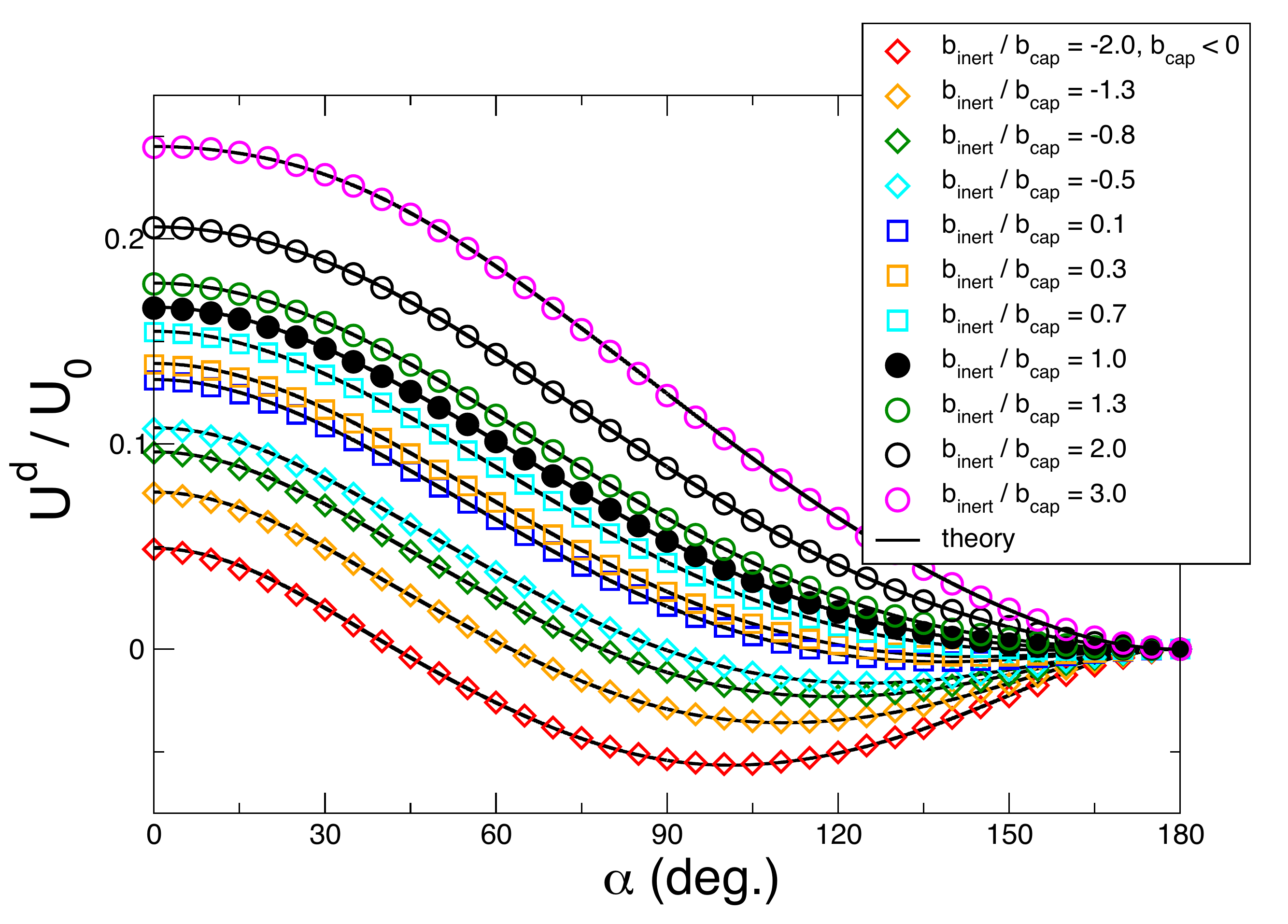}
\caption{\label{fig:ud} Velocity of the particle along the direction $\mathbf{\hat{d}}$ as a function of angle $\alpha$ for various surface mobility contrasts. Notably, for a surface mobility contrast $b_{inert}/b_{cap}$ less than a critical value approximately equal to $b_{inert}/b_{cap} = -0.5$, the particle changes its ``inert-forward'' character of motion ($U^{d} > 0$) to a ``cap-forward'' character ($U^{d} < 0$) for a range of angles $\alpha$. The theoretical curves were obtained by truncating Eq. (\ref{eq:ud}) at sixth order. Points represented by symbols were obtained numerically.}
\end{figure}

\subsubsection{Translation perpendicular to the particle axis}
For translation perpendicular to the particle axis,
\begin{multline}
U^{p} = - \frac{1}{4 \pi R^{2}} \int \left[ v_{s,\theta} (\hat{x} \cdot \hat{\theta} ) + v_{s,\phi} (\hat{x} \cdot \hat{\phi}) \right] dS \\ = - \frac{1}{4 \pi R^{2}} \int (v_{s,\theta} \cos \theta \cos \phi - v_{s,\phi} \sin \phi ) \, dS.
\end{multline}
We consider the contribution from $v_{s,\theta}$ first:
\begin{multline}
 -\frac{\kappa b_0}{4 \pi D} \,  \sum_{l = 1}^{\infty} \sum_{m = 0}^{l}  \int d{\theta} \, \sin \theta \, \int d{\phi} \, g(\theta) \, \frac{A_{lm}}{l+1} \, \\
 \quad \quad \frac{1}{2} \left[(l + m)(l - m + 1) P_{l}^{m-1}(\cos \theta) - \right. \\
  \left. \quad P_{l}^{m+1}(\cos \theta) \right] \, \cos(m \phi) (\cos \theta) \cos(\phi)  \\
 = -\frac{\kappa b_0}{4 D}  \,  \sum_{l = 1}^{\infty} \frac{A_{l1}}{l+1} \,  \int d{\theta} \, \sin \theta \, g(\theta) \, \frac{1}{2} \left[l (l + 1)  P_{l}^{0}(\cos \theta)  \right. \\ 
 \left. -  P_{l}^{2}(\cos \theta) \right]  \, P_{1}^{0}(\cos \theta) \\ 
= -\frac{\kappa b_0}{4 D}  \,  \sum_{l = 1}^{\infty} \frac{A_{l1}}{l+1} \, \left[l (l + 1) {\cal I}_{l,1}^{0,0} - {\cal I}_{l,1}^{2,0} \right].
\end{multline}

And now we consider the contribution from $v_{s,\phi}$:
\begin{multline}
\frac{\kappa b_0}{4 \pi D} \, \sum_{l = 1}^{\infty} \sum_{m = 0}^{l} \int d{\theta} \, \sin \theta \, \int d{\phi} \, g(\theta) \, \frac{m  A_{lm}}{l+1} \, \frac{1}{\sin \theta} \\ \, P_{l}^{m}(\cos \theta) \, \sin(m \phi) \sin(\phi) \\
\begin{split}
&= \frac{\kappa b_0}{4 D} \, \sum_{l = 1}^{\infty} \int d{\theta} \, g(\theta) \, \frac{A_{l1}}{l+1} \, P_{l}^{1}(\cos \theta) \\
&= \frac{\kappa b_0}{4 D} \, \sum_{l = 1}^{\infty}  \, \frac{A_{l1}}{l+1} \, \int d{\theta} \, g(\theta) P_{l}^{1}(\cos \theta) \\
&= \frac{\kappa b_0}{4 D} \, \sum_{l = 1}^{\infty}  \, \frac{A_{l1}}{l+1} \, 2 \, {\cal M}_{l,0}^{1,0},
\end{split}
\end{multline}
where we have defined
\begin{equation}
{\cal M}_{l,k}^{m,n} \equiv \frac{1}{2} \, \int d{\theta} \, g(\theta) P_{l}^{m}(\cos \theta) P_{k}^{n}(\cos \theta).
\end{equation}
Putting the two contributions together, we obtain
\begin{equation}
\label{eq:ux}
U^{p} = \frac{\kappa b_0}{4D} \, \sum_{l=1}^{\infty}  \, \frac{A_{l1}}{l+1} \left[ 2 \, {\cal M}_{l,0}^{1,0} - l (l + 1) {\cal I}_{l,1}^{0,0} + {\cal I}_{l,1}^{2,0} \right]
\end{equation}

\begin{figure}[htb]
\includegraphics[width=\columnwidth]{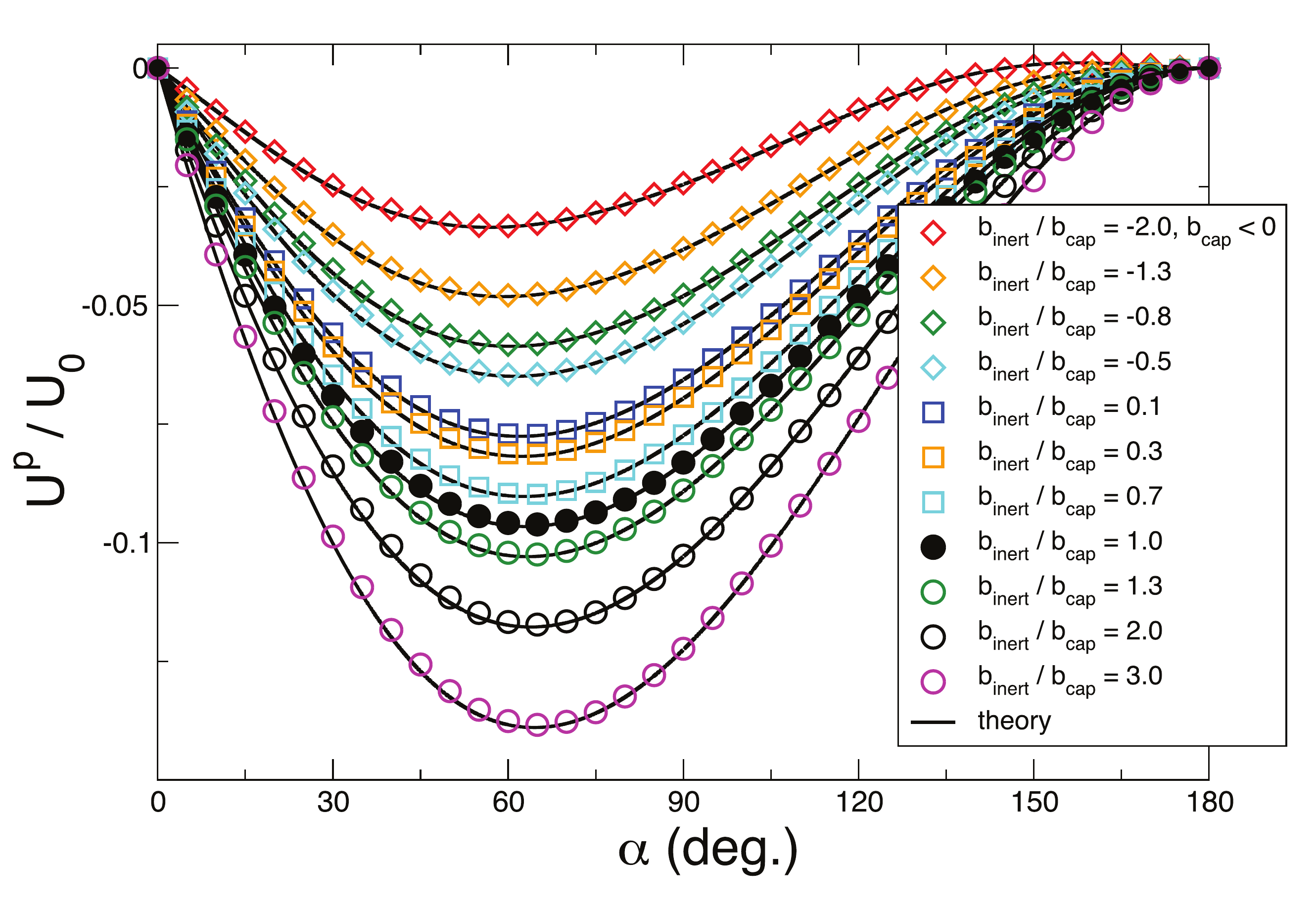}
\caption{\label{fig:up} Velocity of the particle perpendicular to the particle axis, i.e., along the direction $\hat{x}$ defined in the co-moving frame, as a function of angle $\alpha$ for various surface mobility contrasts. The theoretical curves were obtained by truncating Eq. (\ref{eq:ux}) at fifth order.  Points represented by open symbols were obtained numerically.}
\end{figure}

\subsubsection{Rotation}
For rotation of the particle,
\begin{align}
\Omega_{y} &= - \frac{3}{8 \pi R^{3}} \int \hat{y} \cdot (\mathbf{n} \times \mathbf{v}_{s}) \, dS, \\
 &= - \frac{3}{8 \pi R^{3}} \int \hat{y} \cdot (v_{s,\theta} \hat{\phi} - v_{s,\phi} \hat{\theta}) \, dS, \nonumber \\
&= - \frac{3}{8 \pi R^{3}}  \int (v_{s,\theta} \cos \phi - v_{s,\phi} \cos \theta \sin \phi) \, dS. \nonumber
\end{align}
We consider the contribution from $v_{s,\theta}$ first:
\begin{multline}
-  \frac{3 \kappa b_0}{8 \pi R D} \, \int d{\theta} \, \sin \theta \, \int d{\phi} \, g(\theta) \,  \sum_{l = 1}^{\infty} \sum_{m = 0}^{l}  \frac{A_{lm}}{l+1} \, \\
  \frac{1}{2} \left[(l + m)(l - m + 1) P_{l}^{m-1}(\cos \theta) \right. \\ 
\left. - P_{l}^{m+1}(\cos \theta) \right]  \, \cos(m \phi)  \cos \phi  \\
= - \frac{3 \kappa b_0}{8 R D} \, \int d{\theta} \, \sin \theta \, g(\theta) \,  \sum_{l = 1}^{\infty}  \frac{A_{l1}}{l+1} \,  \frac{1}{2} \left[l (l + 1)  P_{l}^{0}(\cos \theta)  \right. \\ \left. - P_{l}^{2}(\cos \theta) \right]  \\
= - \frac{3 \kappa b_0}{8 R D} \,  \sum_{l = 1}^{\infty}  \frac{A_{l1}}{l+1} \,  \int d{\theta} \, \sin \theta \, g(\theta) \, \frac{1}{2} \left[l (l + 1)  P_{l}^{0}(\cos \theta) \right. \\ \left. - P_{l}^{2}(\cos \theta) \right]   \\
=  - \frac{3 \kappa b_0}{8 R D} \,  \sum_{l = 1}^{\infty}  \frac{A_{l1}}{l+1} \, \left[ l (l + 1) {\cal I}_{l,0}^{0,0} - {\cal I}_{l,0}^{2,0} \right].
\end{multline}
And now we consider the contribution from $v_{s,\phi}$:
\begin{multline}
\frac{3 \kappa b_0}{8 \pi R D}  \sum_{l = 1}^{\infty} \sum_{m = 0}^{l} \int d{\theta} \, \sin \theta \, \int d{\phi} \, g(\theta) \, \\ \frac{m  A_{lm}}{l+1} \, \frac{1}{\sin \theta} \, P_{l}^{m}(\cos \theta) \, \sin(m \phi) \, \cos \theta  \, \sin \phi \\ 
\begin{split}
&= \frac{3 \kappa b_0}{8 R D}  \sum_{l = 1}^{\infty} \int d{\theta} \, g(\theta) \, \frac{A_{l1}}{l+1} \,  P_{l}^{1}(\cos \theta) \, \cos(\theta)  \\
&= \frac{3 \kappa b_0}{4 R D} \sum_{l = 1}^{\infty}  \, \frac{A_{l1}}{l+1} {\cal M}_{l,1}^{1,0}.
\end{split}
\end{multline}
Putting everything together,
\begin{equation}
\label{eq:omega_y}
\Omega_{y} = \frac{3 \kappa b_0}{8 R D}  \sum_{l = 1}^{\infty}  \, \frac{A_{l1}}{l+1} \left[2 \, {\cal M}_{l,1}^{1,0} - l (l + 1)  {\cal I}_{l,0}^{0,0} + {\cal I}_{l,0}^{2,0}  \right].
\end{equation}

\begin{figure}[htb]
\includegraphics[width=\columnwidth]{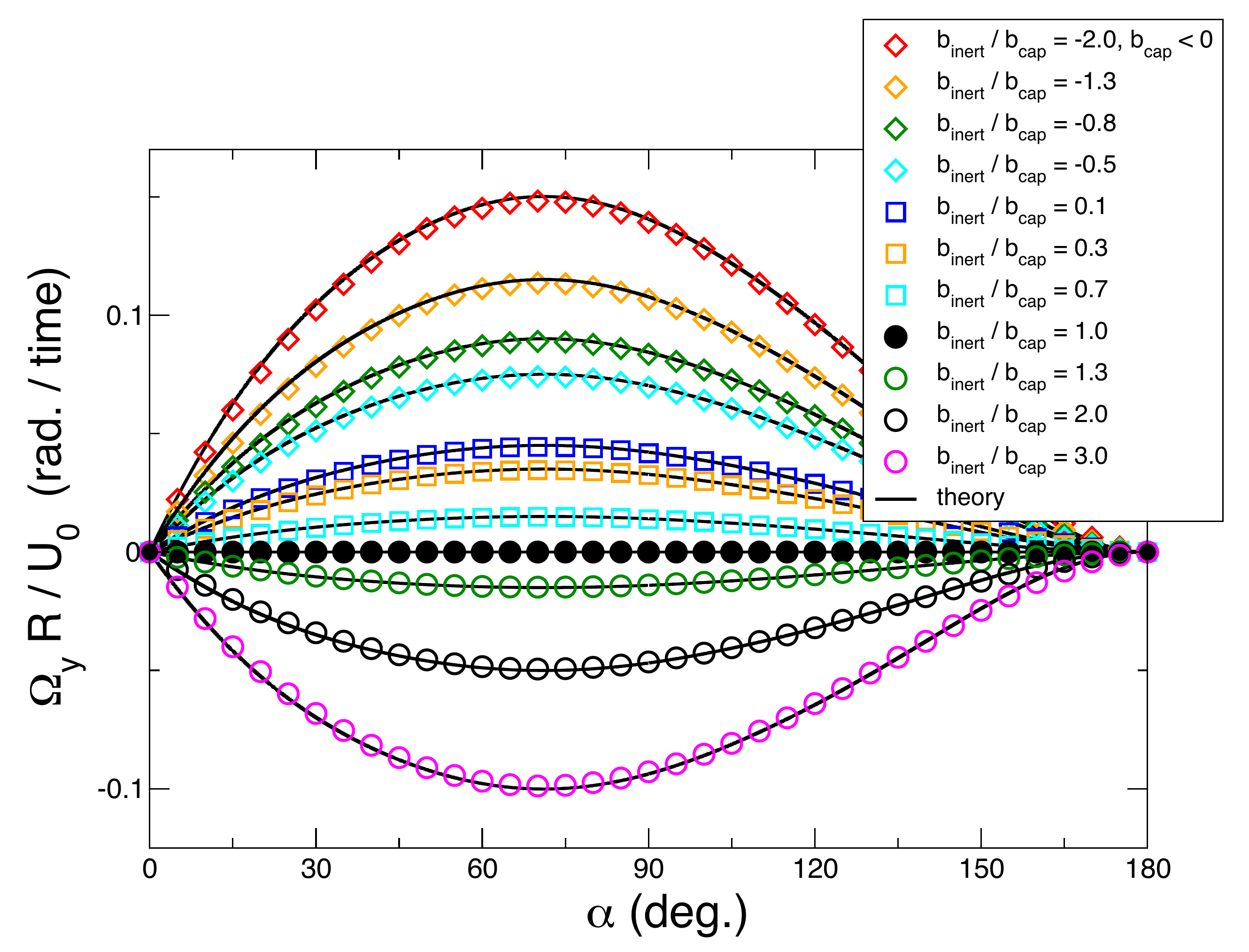}
\caption{\label{fig:DalphaDt} Rotational velocity $\Omega_{y}$ as a function of angle $\alpha$ for various surface mobility contrasts and $b_{cap} < 0$. For $b_{inert}/b_{cap} > 1$, the particle rotates its catalytic cap to face the light; otherwise, the particle rotates its inert face to face the light. The theoretical curves were obtained by truncating Eq. (\ref{eq:omega_y}) at fifth order.  Points represented by open symbols were obtained numerically.}
\end{figure}

\subsubsection{Application to particle with piecewise constant surface mobility}

In order to evaluate Eqs. (\ref{eq:ud}), (\ref{eq:ux}), and (\ref{eq:ud}), we must specify the function $b_0 \, g(\theta)$. A reasonable choice would be to assume that the phoretic mobility is a constant over the surface of the cap, $b_{cap}$, and a potentially different constant over the inert face, $b_{inert}$. Further, we choose $b_0 = b_{cap}$, define the surface mobility ratio $\tilde{b} \equiv b_{inert}/b_{cap}$, and define the characteristic velocity $U_0$ as $U_0 \equiv | {b_{cap} \kappa}/{D}|$.
We obtain:
\begin{multline}
\label{eq:ud2}
U^{d} = U_0 \, \sgn(b_{cap}) \, \left[ \frac{1}{6} (1 + \tilde{b}) A_{10} - \frac{1}{8} (1 - \tilde{b}) A_{20}  \right. \\ \left. + \frac{1}{24} (1 - \tilde{b}) A_{40} + ... \right],
\end{multline}
\begin{multline}
\label{eq:ux2}
U^{p} = \frac{1}{4} \, U_0 \, \sgn(b_{cap}) \, \left[ -\frac{2}{3} (1 + \tilde{b}) A_{11} + \frac{1}{2} (1 - \tilde{b}) A_{21}  \right. \\ 
\left.  - \frac{1}{6} (1 - \tilde{b}) A_{41} +  ... \right],
\end{multline} 
\begin{multline}
\label{eq:omega_y2}
\Omega_{y} = \frac{3}{8} \frac{U_0}{R }  \, \sgn(b_{cap}) \, (1 - \tilde{b}) \left[- \frac{1}{2} A_{11} + \frac{3}{8} A_{31} \right. \\ \left. - \frac{5}{16} A_{51} + ...  \right].
\end{multline}
If the cap faces the light ($\alpha = 0^{\circ}$), we can obtain a highly accurate approximate expression for $U^{d}$ by evaluating Eq. (\ref{eq:ud2}) up to $l = 6$:
\begin{multline}
\label{eq:Ud_alpha_0}
\frac{U^{d}(\alpha = 0^{\circ})}{U_{0}} \approx - \frac{\sgn(b_{cap})}{12} \left(1 + \tilde{b} \right) - \\  \sgn(b_{cap}) \frac{1447}{32768} \left(1 - \tilde{b} \right).
\end{multline}

Eqs. (\ref{eq:ud2}), (\ref{eq:ux2}), and (\ref{eq:omega_y2}) exhibit excellent agreement with numerical calculations, obtained with the boundary element method, for various choices of the parameter $\tilde{b}$. The numerical data and analytical calculations are shown in Figs. \ref{fig:ud}, \ref{fig:up}, and \ref{fig:DalphaDt}.  

\begin{figure}[htb]
\includegraphics[width=\columnwidth]{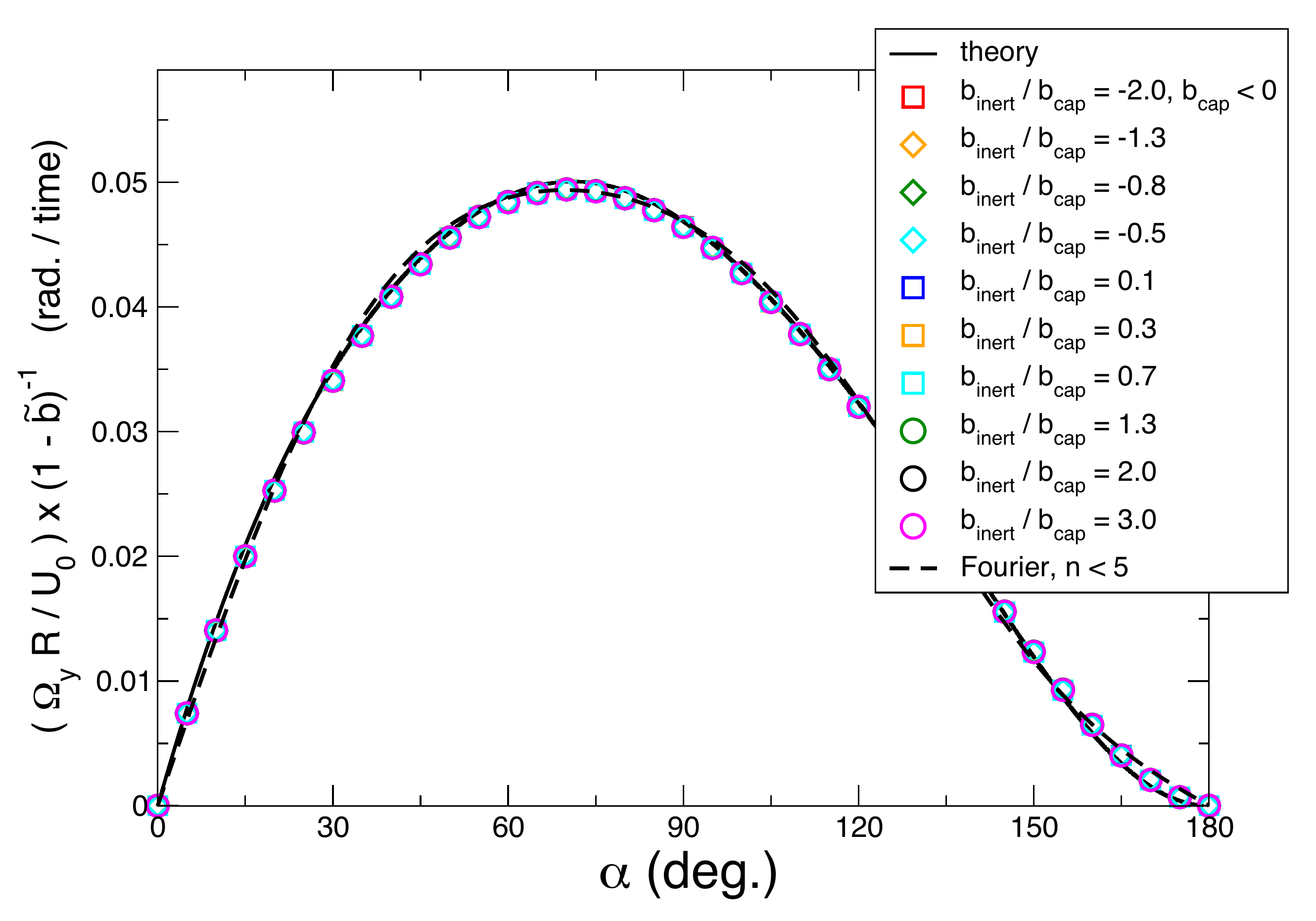}
\caption{\label{fig:rotation_collapsed}  Rotational velocity as a function of angle $\alpha$ for \textcolor{black}{$b_{cap} < 0$} and various values of the mobility contrast $\tilde{b}$, collapsed onto a single master curve upon rescaling by a factor of $(1 - \tilde{b})$. Additionally, we show the Fourier representation of Eq. (\ref{eq:fourier}), truncated to $n = 4$. }
\end{figure}

It is evident from inspection of Eq. (\ref{eq:omega_y2}) and Fig. \ref{fig:DalphaDt} that the curves in Fig. \ref{fig:DalphaDt} can be collapsed onto one master curve by rescaling the data by a factor of $\sgn(b_{cap})\,(1 - \tilde{b})$. This collapse is shown in Fig. \ref{fig:rotation_collapsed}. It is clear that master curve is approximately, but not quite, a sinusoid. Therefore, one expects that it can be reasonably approximated by truncating a Fourier expansion
\begin{equation}
\label{eq:fourier_1}
\frac{\textcolor{black}{\Omega_y} R}{U_{0}} = \sgn(b_{cap}) (1 - \tilde{b}) \sum_{n = 1}^{\infty} a_{n} \sin(n \alpha)
\end{equation}
after a few terms. Analytically, we calculate 
\begin{equation}
\label{eq:fourier}
a_n = \sgn(b_{cap}) (1 - \tilde{b})^{-1} \frac{2}{\pi} \frac{R}{U_0} \int_{0}^{\pi} \Omega_y(\alpha) \, \sin(n \alpha) \, d\alpha,
\end{equation}
obtaining $a_{1} = 3/64$, $a_{2} = 1531/13440 \pi^{2}$, $a_{3} = 0$, and $a_{4} = 703/33600 \pi^{2}$. Truncating the Fourier series to $n < 5$, we obtain a reasonable approximation to the theoretical master curve and the numerical data. The Fourier series representation is of interest for at least two reasons. First, it may be less cumbersome than the analytical expressions for $A_{lm}$ for use in calculations. Secondly, it provides additional insight into the physical properties of phototactic Janus particles. We see that the first order term, $\sin \alpha$, drives polar alignment with respect to $\mathbf{\hat{q}}$. However, there are also higher order terms, such as $\sin 2 \alpha$, that have nematic symmetry. For comparison, rotation due to bottom-heaviness is strictly polar, with $\Omega_{y}^{bh} \sim \sin \theta_g$ (where $\theta_g$ is the orientation of $\mathbf{\hat{d}}$ with respect to the vertical.)

Naturally, it is interesting to consider the components of the translational velocity of the particle in a stationary frame. We define the ``primed'' frame to have $\hat{z}'$ in the $\mathbf{\hat{q}}$ direction. Accordingly, the particle orientation vector $\mathbf{\hat{d}}$ has an angle $\alpha$ with respect to $\hat{z}'$.  The projection of $\mathbf{\hat{d}}$ in the plane spanned by $\hat{x}'$ and $\hat{y}'$ has an angle $\phi'$ with respect to $\hat{x}'$. 

\begin{figure}[htb]
\includegraphics[width=\columnwidth]{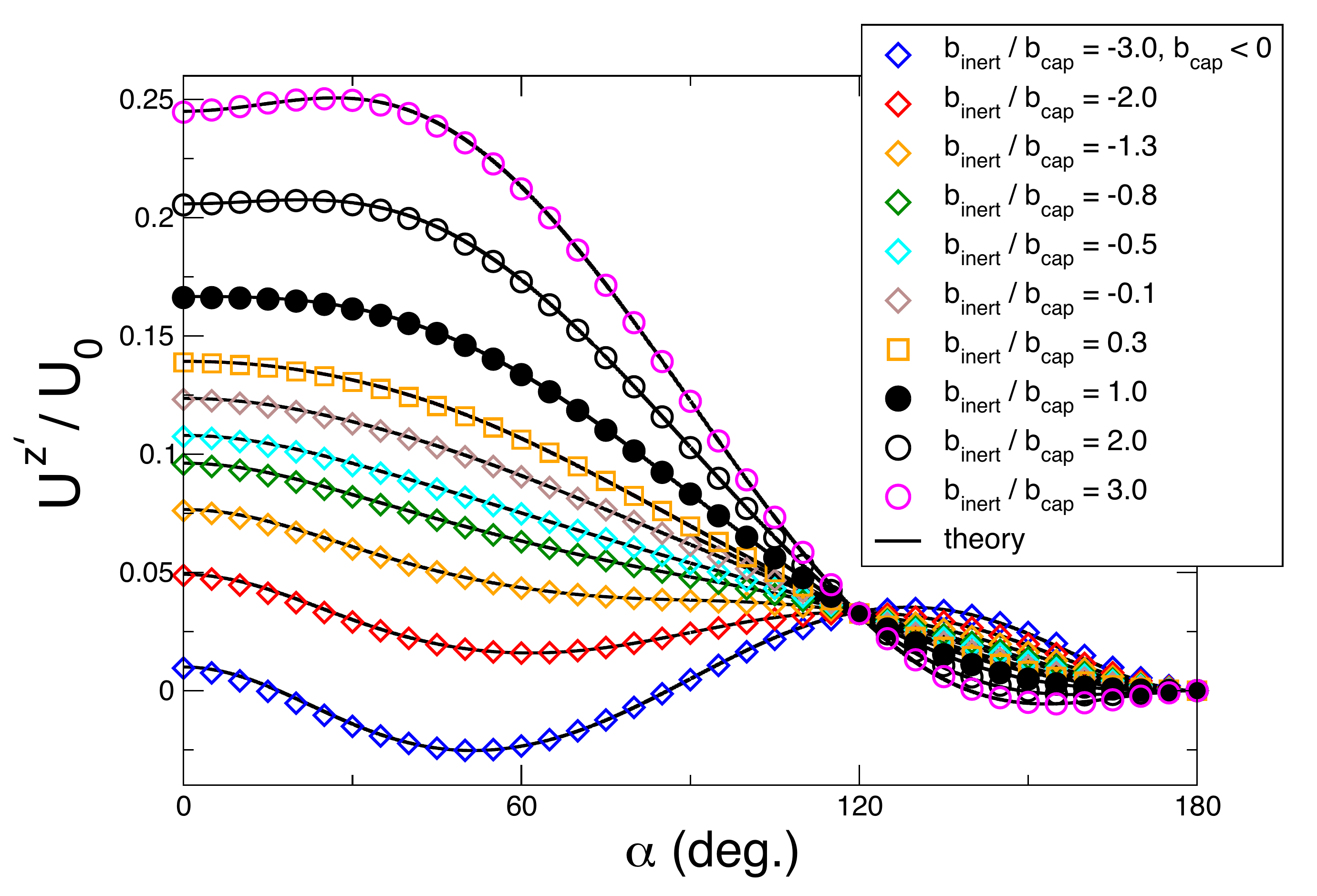}
\caption{\label{fig:Uz_lab} Component of the translational velocity of the particle in the $\hat{z}'$ direction of the stationary frame, where $\hat{z}' = \mathbf{\hat{q}}$, as a function of orientation $\alpha$.}
\end{figure}

\begin{figure}[htb]
\includegraphics[width=\columnwidth]{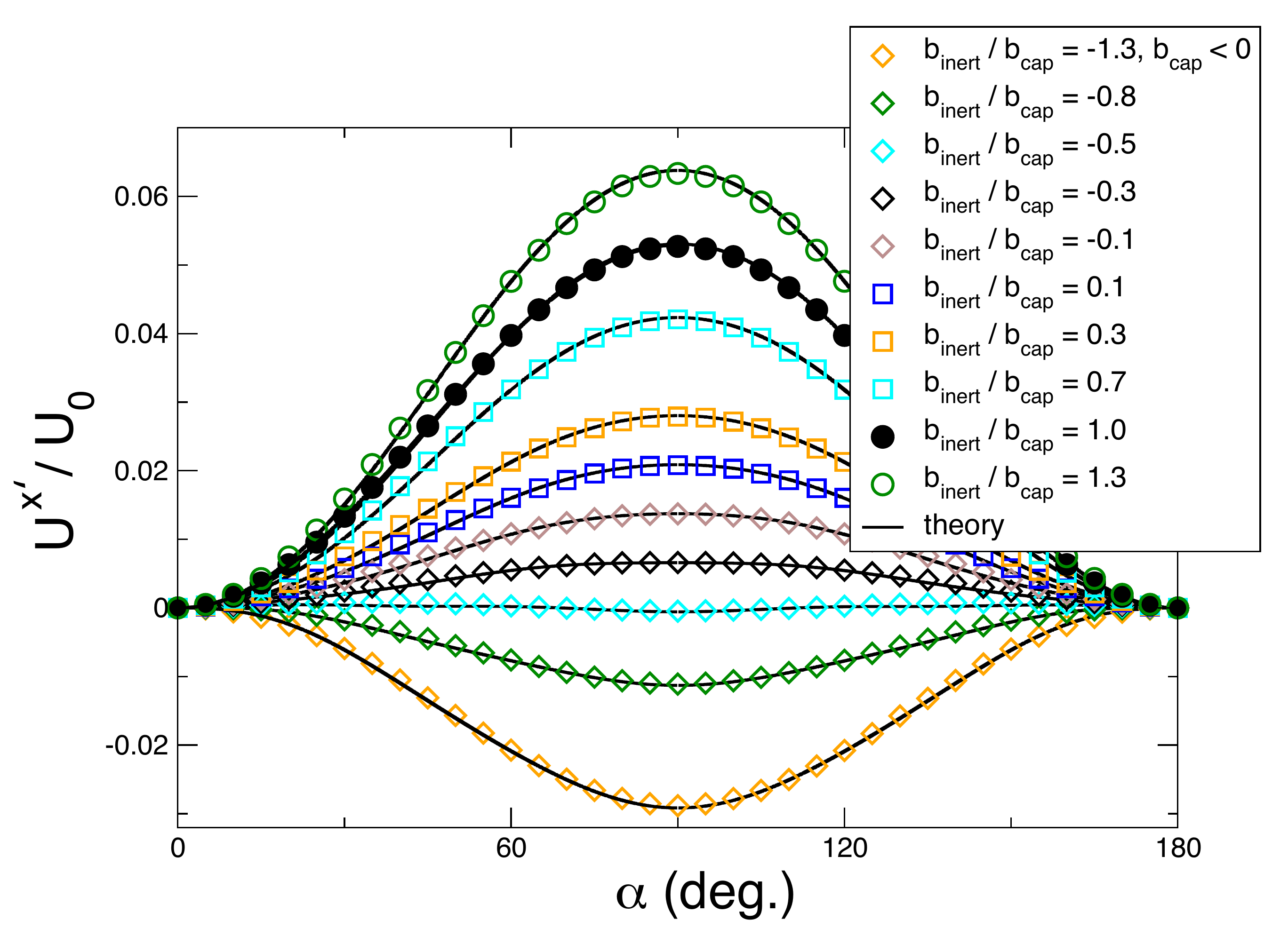}
\caption{\label{fig:Ux_lab} Component of the translational velocity of the particle \textcolor{black}{in the} $\hat{x}'$ direction of the stationary frame as a function of orientation $\alpha$ for $\phi' = 0^{\circ}$. }
\end{figure}

Having defined the stationary coordinate system, we can transform the particle \textcolor{black}{velocity} in the stationary frame:
\begin{equation}
\label{eq:Uxprime}
U^{x'} = U^{d} \sin(\alpha) \cos(\phi') + U^{p} \cos(\alpha) \cos(\phi')
\end{equation}
\begin{equation}
\label{eq:Uyprime}
U^{y'} = U^{d} \sin(\alpha) \sin(\phi') + U^{p} \cos(\alpha) \sin(\phi')
\end{equation}
\begin{equation}
\label{eq:Uzprime}
U^{z'} = U^{d} \cos(\alpha) - U^{p} \sin(\alpha)
\end{equation}
In Fig. \ref{fig:Uz_lab}, we show $U^{z'}$ as a function of the angle $\alpha$ for different values of $\tilde{b}$. Interestingly, for the orientation $\alpha = 120^{\circ}$, there is no dependence of $U^{z'}$ on the the surface mobility contrast $\tilde{b}$. Since only the surface mobility of the cap is involved in the definition of the velocity scale $U_0$,  this independence of $\tilde{b}$ indicates that inert region of the particle has no net contribution to $U^{z'}$ at $\alpha = 120^{\circ}$. \textcolor{black}{We also consider the component of the particle velocity in the $x'y'$ plane. Since the system is rotationally symmetric around the $z'$ axis, we take $\phi' = 0^{\circ}$ without loss of generality and show $U^{x'}$ as a function of $\alpha$ in Fig. \ref{fig:Ux_lab}.}

\subsection{Effect of level of coverage by catalyst}

\begin{figure}[htb]
\includegraphics[width=\columnwidth]{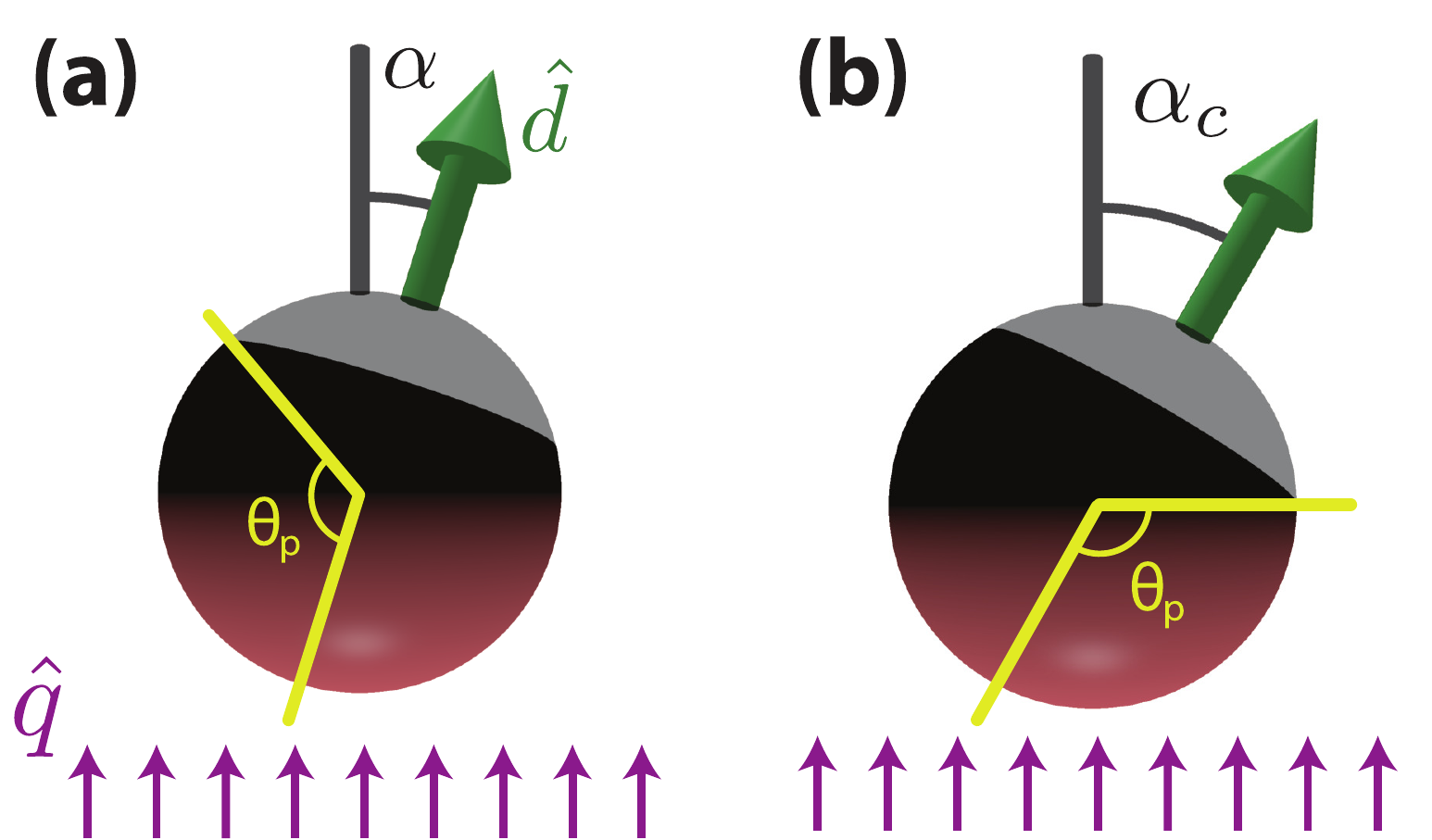}
\caption{\label{fig:large_cap} (a) Schematic illustration of a particle with a large catalytic cap. The cap size is characterized by the cap opening angle $\theta_p$, where $\theta_p = 90^{\circ}$ for a particle that is half covered by catalyst. (b) Schematic illustration of the critical angle $\alpha_{c} \equiv \theta_p - 90^{\circ}$ for a particle with a large cap. For $0 \leq \alpha \leq \alpha_{c}$, the area of illuminated catalyst is exactly half of the total area of the particle surface, and does not depend on $\alpha$.  For $\alpha > \alpha_{c}$, some of the inert region of the particle surface is exposed to light, and hence the area of illuminated catalyst changes with $\alpha$. }
\end{figure}

All of the preceding analysis considered a particle that is half covered by catalyst. Here, we consider an arbitrary level of coverage, parameterized by the opening angle $\theta_p$ of the catalytic cap (Fig. \ref{fig:large_cap}(a)).  Our analytical framework can be extended to arbitrary $\theta_p$, but the resulting expressions involve hypergeometric series, and hence do not provide much physical insight. Therefore, we restrict ourselves in this section to numerical calculations.

\begin{figure}[htb]
\includegraphics[width=\columnwidth]{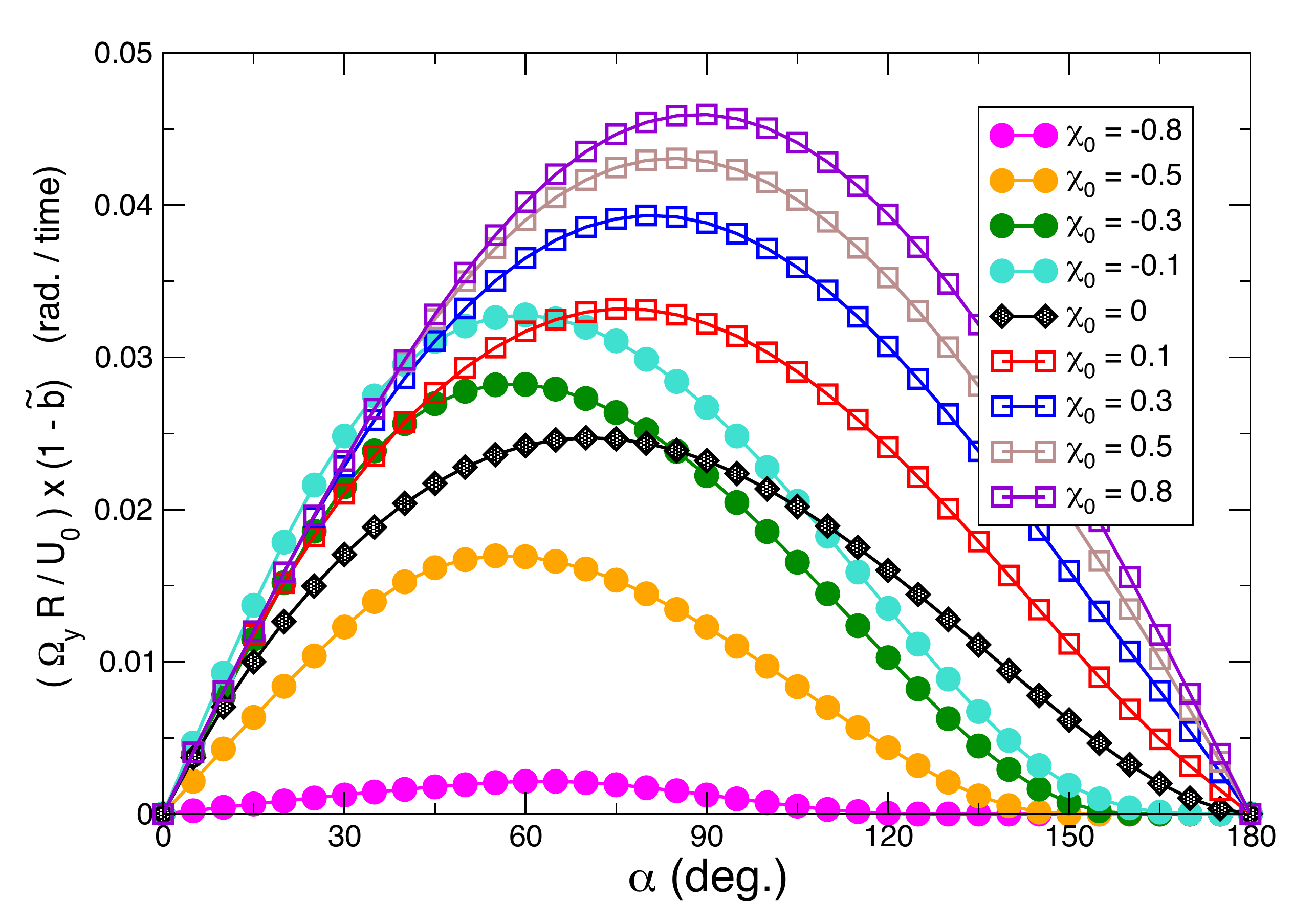}
\caption{\label{fig:rotation_coverage} Rotational velocity \textcolor{black}{$\Omega_{y}$} as a function of angle $\alpha$ for various values $\chi_{0}$ of the level of coverage by catalyst. The data has been numerically obtained by the BEM and scaled by $(1 - \tilde{b})$ in order to obtain master curves, as described in the text. Lines are shown to guide the eye. }
\end{figure}

We define the coverage parameter $\chi_0 \equiv -\cos(\theta_p)$. This parameter ranges between $\chi_0 = -1$ for no coverage and $\chi_0 = 1$ for a completely covered particle, with $\chi_0 = 0$ for a half covered particle. In Fig. \ref{fig:rotation_coverage}, we show $d{\alpha}/dt$ as a function of $\alpha$ for various values of $\chi_0$. For each value of $\chi_0$, curves obtained for various values of $\tilde{b}$ collapse onto a master curve when rescaled by $(1 - \tilde{b})$.  

The family of master curves parameterized by $\chi_0$ has some interesting features. As $\chi_0$ is increased from half coverage ($\chi_0 = 0$), the master curves become more and more sinusoidal, with some slight deviation occuring near $\alpha = 180^{\circ}$. The explanation of this is following: within the stationary reference frame, we consider a particle with a large cap (Fig. \ref{fig:large_cap}(a)). We see that for small angles $\alpha$, the region of the particle surface with $90^{\circ} < \theta < 180^{\circ}$ is catalytic and fully illuminated. This occurs for a broad range of angles $0 \leq \alpha \leq \alpha_c$, where $\alpha_{c} = \theta_p - 90^{\circ}$ (Fig. \ref{fig:large_cap}(b)). For angles greater than the critical angle, $\alpha > \alpha_c$, a region of the inert face of the particle is illuminated; therefore, the area of illuminated catalyst has decreased. The area of illuminated catalyst decreases with $\alpha$ for $\alpha > \alpha_c$. Therefore, we can distinguish two regimes: \newline
(i) For $0 \leq \alpha \leq \alpha_c$, the solute number density field around the particle does not change with $\alpha$. The particle can be regarded as being in an effectively ``external'' or fixed solute gradient. It is known that for a Janus particle in an externally maintained concentration gradient, the rate of rotation varies as $\Omega_y \sim (1 - \tilde{b}) \sin(\alpha)$.\cite{anderson84} \newline
(ii) For $\alpha_c < \alpha \leq 180^{\circ}$, the area of illuminated catalyst changes with $\alpha$. This shadowing effect induces phototactic rotation.

\begin{figure}[htb]
\includegraphics[width=\columnwidth]{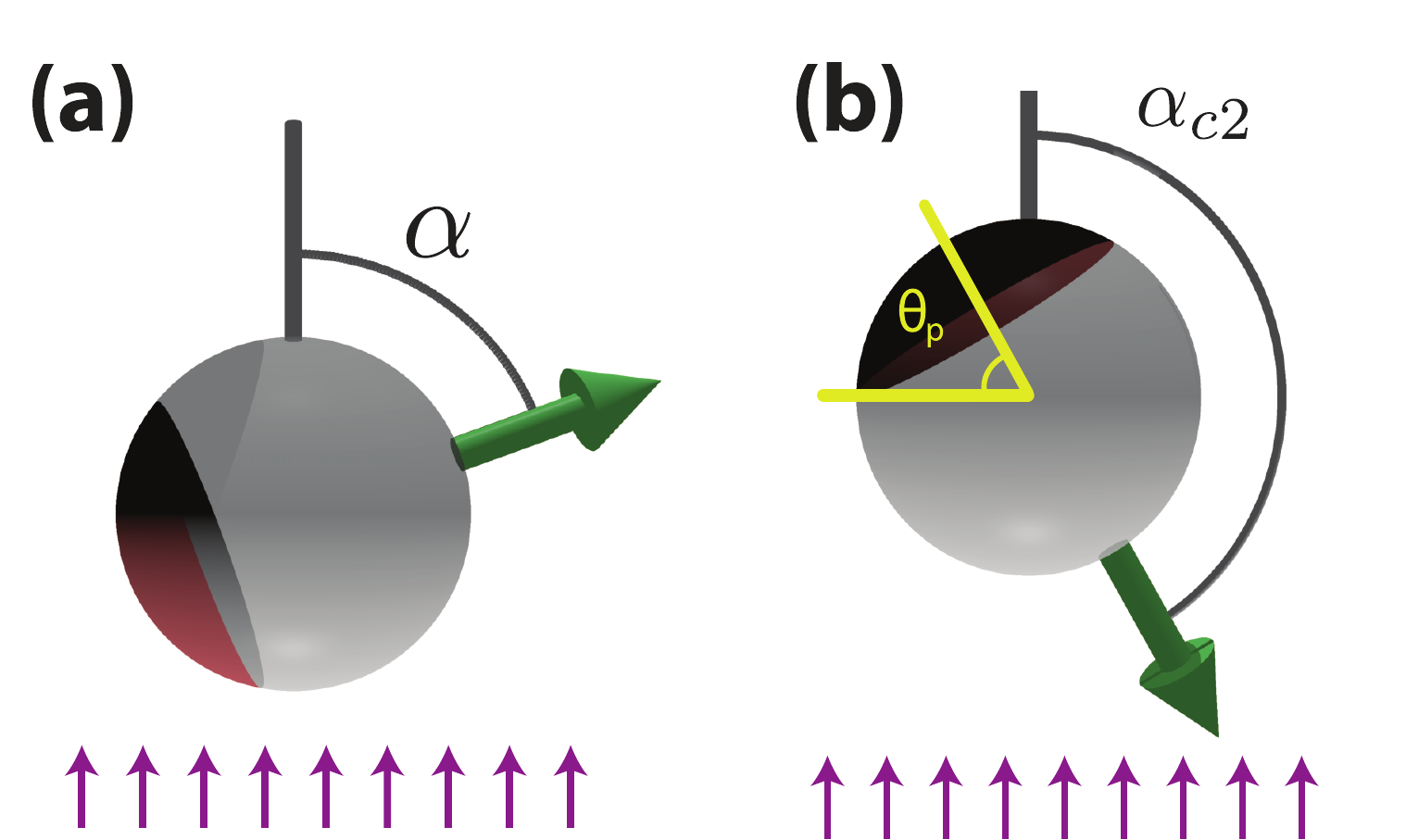}
\caption{\label{fig:small_cap} (a) Schematic illustration of a particle with a small cap.  (b) Illustration of the critical angle $\alpha_{c2} \equiv 90^{\circ} + \theta_p$. For $\alpha > \alpha_{c2}$, the catalytic cap is completely in shadow, and the particle is inactive.  }
\end{figure}

Now we consider the effect of decreasing the cap size from half coverage, i.e., we consider the range of values $-1 \leq \chi_0 < 0$. We find that for particles with small caps, there is a certain range of angles $\alpha$ for which $\Omega_y$ is exactly zero. This range increases in width as $\chi_0$ decreases. This phenomenon occurs because, for small caps, there is a broad range of angles $\alpha_{c2} < \alpha \leq 180^{\circ}$, where $\alpha_{c2} = 90^{\circ} + \theta_p$, for which the cap is completely in shadow (Fig. \ref{fig:small_cap}(b)). We recall that, for half coverage, full shadowing of the cap occurs only at the value $\alpha = 180^{\circ}$, which is captured by the expression for $\alpha_{c2}$. Furthermore, for large caps ($\chi_0 > 0$), $\alpha_{c2}$ takes unphysical values $\alpha_{c2} > 180^{\circ}$, and hence there is no angle $\alpha$ for which the cap is completely in shadow.  

For low coverage particles, \textcolor{black}{$\Omega_y$} shows significant deviation from a sinsuoidal form when considered over the whole range $\alpha \in [0, \pi]$. However, within the range $\alpha \in [0, \alpha_{c2}]$, the function appears very much like a sinusoid with a period $\alpha_{ac2}$. Accordingly, we can approximate the function as the following:

\begin{multline}
\label{eq:fourier_low}
 \frac{\textcolor{black}{\Omega_y} R}{U_0} = \sgn(b_{cap}) (1 	- \tilde{b}) \, \Theta(\alpha_{ac2} - \alpha) \\ \sum_{n = 1}^{\infty} a_{n} \sin\left( \frac{n \, \pi \, \alpha }{\alpha_{ac2}}\right),
\end{multline}
with the coefficients determined by numerical fitting. Truncating Eq. (\ref{eq:fourier_low}) to $n < 5$ provides a good approximation to the \textcolor{black}{numerical} data, and will be useful in the next section. The numerically fitted coefficients are provided in Table \ref{table:fourier}.

\begin{table}
\centering
\begin{tabular}{ |c|c|c|c|c|c| } 
 \hline
 $\chi_0$ & $a_1$ & $a_2$ & $a_3$ & $a_4$ & $\alpha_{ac2}$ \\
 \hline
 -0.8 & 0.00375046 & 0.000352037 & -0.0005005 & $0.000152144$ & $126.9^{\circ}$  \\ 
 -0.5 & 0.0300142 & 0.00824731 & -0.00112485 & 0.000247047 & $150^{\circ}$ \\ 
 -0.3 & 0.0530889 & 0.0114933 & 4.51E-05 & 0.000891105 & $162.5^{\circ}$ \\ 
 -0.1 & 0.0553248 & 0.0222223 & -1.99E-04 & 0.00200734 & $174.2^{\circ}$ \\ 
 \hline
\end{tabular}
 \caption{Table of numerically fitted coefficients in a truncated Fourier expansion (Eq. \ref{eq:fourier_low} with $n < 5$) for various coverage levels $\chi_0 < 0$.}
 \label{table:fourier}
\end{table}

\section{Fluctuating photoactive particle \label{Sec:Fluctuating}}

The preceding analysis considered the deterministic contributions of self-phoresis to particle motion. In the following, we also consider the effect of thermal noise on the motion of the particle. In particular, we  study the interplay of phototactic or anti-phototactic alignment with rotational Brownian motion.  This interplay determines the probability distribution of the particle orientation, and therefore the long-time behavior of the particle.  In addition, we will also consider the contributions of gravity to motion of a particle. Generally, catalytic Janus particles are heavier than water and, in the absence of chemical activity, will sediment in solution; therefore, for an active particle to migrate vertically, its average vertical swimming velocity must exceed the sedimentation velocity $V^s$. Additionally, catalytic Janus particles are typically bottom-heavy, owing to their ``patchy'' catalytic coating, and tend to align cap-down with the vertical. Bottom-heaviness can compete with or enhance phototactic or anti-phototactic alignment, depending on the direction of illumination.

In the preceding analysis, we designated the self-phoretic velocities by $\mathbf{U}$ and $\bm{\Omega}$. In the following, to more clearly the distinguish self-phoretic and gravitational contributions to particle velocity, we designate the self-phoretic velocities by $\mathbf{U}_{swim}$ and $\bm{\Omega}_{swim}$. 


\subsection{Effective potential for a half-covered particle}

It possible to incorporate the effect of thermal noise by mapping our findings into the framework of equilibrium statistical mechanics. Consider that the rate of rotation \textcolor{black}{$\Omega_y$} is a single-valued function of $\alpha$ (and not of spatial position). Examining Fig. \ref{fig:DalphaDt}, it is clear that each of these curves is a derivative of some function $U_{eff}(\alpha)$:
\begin{equation}
\label{eq:U_eff_1}
\textcolor{black}{{\Omega}_y} = - \frac{1}{8 \pi \mu R^{3}} \frac{d U_{eff}}{d \alpha},
\end{equation}
where we have included the rotational drag coefficient as a prefactor. Accordingly, we define $U_{eff}$ for a half-covered particle as the following:
\begin{equation}
\label{eq:ueff_defn}
U_{eff}(\alpha) \equiv - 8 \pi \mu R^{3} \int_{\pi}^{\alpha} \Omega_{y}(\alpha') \, d\alpha'.
\end{equation}
We define the dimensionless effective potential as 
\begin{equation}
\label{eq:U_eff_2}
\tilde{U}_{eff} \equiv \frac{R}{V^{s}} \, \frac{U_{eff}}{8 \pi \mu R^{3}}.
\end{equation}
Here, we have assumed that the particle has a non-zero sedimentation velocity $V^{s}$, and chosen it as a characteristic velocity scale. If we substitute the Fourier representation of Eq. (\ref{eq:fourier_1}), we obtain
\begin{multline}
\label{eq:tilde_U_eff}
\tilde{U}_{eff}(\alpha) = \sgn(b_{cap}) \, \frac{U_0}{V^s} \, (1 - \tilde{b})  \\  \sum_{n \geq 1} \frac{a_{n}} {n} \left[ \cos(n \, \alpha) - (-1)^n \right]
\end{multline}
for the half-covered particle.


\textcolor{black}{In order to obtain the long-time probability distribution of particle orientations $P(\alpha)$, we write the steady Smoluchowski equation
\begin{equation}
-\beta D_{R} \frac{1}{\sin \alpha} \frac{d}{d \alpha} \left(\dot{\alpha} P(\alpha) \sin \alpha \right) + D_R \frac{1}{\sin \alpha} \frac{d}{d \alpha} \left( \sin \alpha \frac{d P}{d \alpha} \right) = 0.
\end{equation}
Here, $D_R$ is the rotational diffusion coefficient of the particle, and $\beta^{-1} = k_B T$. This equation can be solved to obtain a Boltzmann distribution,}
\begin{equation}
\label{eq:boltzmann_half}
P(\alpha) = \mathcal{N} \exp \left[ -Pe^{r} \, \tilde{U}_{eff}(\alpha) \right].
\end{equation}
Here, $Pe^{r}$ is the rotational P\'{e}clet number of the particle,
\begin{equation}
Pe^{r} \equiv \frac{V^{s}}{R} \frac{8 \pi \mu R^{3}}{k_{B} T},
\end{equation}
and $\mathcal{N}$ is a normalizing prefactor. Note that the rotational P\'{e}clet number is related to the familiar translational P\'{e}clet number of the particle,
\begin{equation}
Pe^{t} \equiv V^{s} {R} \, \frac{6 \pi \mu R}{k_{B} T}
\end{equation}
by $Pe^{r} = \frac{4}{3} Pe^{t}$. 

\begin{figure}[htb]
\includegraphics[width=0.9\columnwidth]{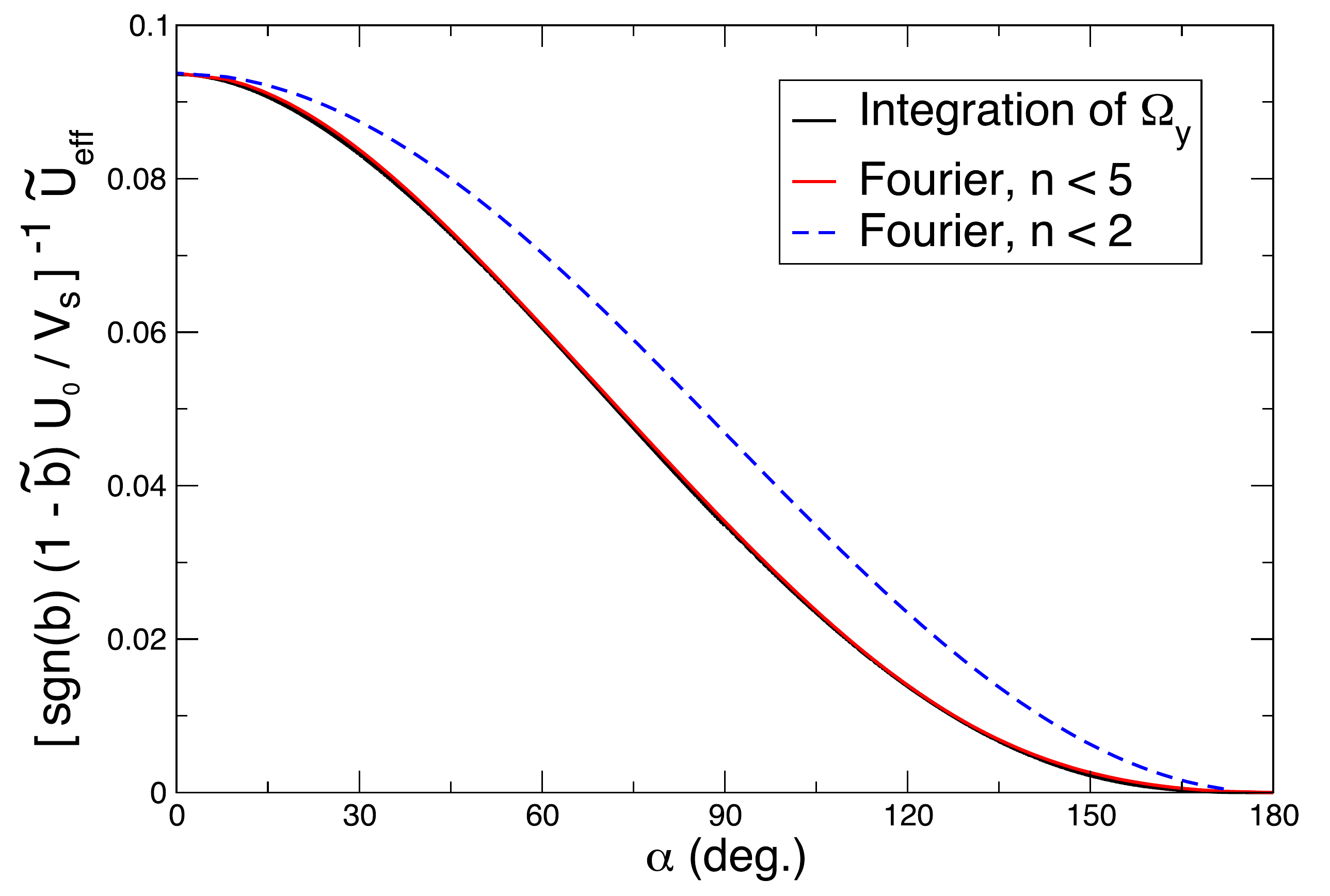}
\caption{\label{fig:eff_pot_no_G} Rescaled effective potential $\tilde{U}_{eff}$ for a half-covered particle. The solid black curve is obtained by numerical integration of Eq. (\ref{eq:ueff_defn}), using Eq. (\ref{eq:omega_y2}) truncated at $n <5$. The solid red line is obtained using Eq. (\ref{eq:tilde_U_eff}) for $n < 5$. For comparison, we also show (dashed blue line) an effective potential obtained using only the first term in the Fourier expansion of Eq. (\ref{eq:tilde_U_eff}).}
\end{figure}

\textcolor{black}{This completes the mapping into the framework of equilibrium statistical mechanics: the probability distribution of particle orientations is governed by a Boltzmann equation with a nonequilibrium effective potential.} Now we consider the \textcolor{black}{the detailed dependence of the} probability distribution \textcolor{black}{on the} particle activity and \textcolor{black}{on} the P\'{e}clet number. We characterize the activity by the parameter $A$:
\begin{equation}
A \equiv \frac{U^{d,max}}{V^{s}}.
\end{equation}
This is the ratio of the maximum value of the upward swimming velocity $U^{d,max} \equiv U^{d}(\alpha = 0^{\circ})$, given by Eq. (\ref{eq:Ud_alpha_0}), to the sedimentation velocity. If $A > 1$, it is possible, for some range of $\alpha$ in the vicinity of $\alpha = 0^{\circ}$, for the particle to swim upward, against the direction of gravity. We can rewrite Eq. (\ref{eq:tilde_U_eff}) as
\begin{multline}
\label{eq:tilde_U_eff_2}
\tilde{U}_{eff}(\alpha) = \sgn(b_{cap}) \, A \: \frac{U_0}{U^{d,max}} \: (1 - \tilde{b}) \\  \sum_{n \geq 1} \frac{a_{n}} {n} \left[ \cos(n \, \alpha) - (-1)^n \right].
\end{multline}
Since $U^{d} \sim U_0$, the term $U_0/U^{d,max}$ is a numerical constant for a given $\tilde{b}$ and $\sgn({b_{cap}})$ (see Eq. (\ref{eq:Ud_alpha_0})). From rewriting $\tilde{U}_{eff}$ in this way and from examining Eq. (\ref{eq:boltzmann_half}), we see that the probability distribution of orientations, for a given choice of surface mobilities, is controlled by $Pe^{r} A$. In Fig. \ref{fig:PDFs_phototactic_halfCov}, we show the orientational distributions for $\tilde{b} = 1.1$ and various values of $Pe^{r} A$.  The effective potential can be collapsed onto one master curve for various levels of photochemical activity and values of the two surface mobilities by rescaling $\tilde{U}_{eff}$ by the terms on the right hand side of Eq. (\ref{eq:tilde_U_eff_2}) that precede the summation symbol. In Fig. \ref{fig:eff_pot_no_G}, we show the  master curve obtained with Eq. (\ref{eq:omega_y2}), truncated at $n < 5$, and Eq. (\ref{eq:ueff_defn}), as well as the master curve obtained with the truncated Fourier representation of Eq. (\ref{eq:tilde_U_eff}); the two curves closely agree. For comparison, an effective potential with $U_{eff} \sim \cos(\alpha)$, i.e., containing only the first term in the Fourier expansion of the angular velocity, is also shown. The level of photochemical activity and the choice of surface mobilities cannot change the functional form of $U_{eff}(\alpha)$.

\begin{figure}[htb]
\includegraphics[width=0.9\columnwidth]{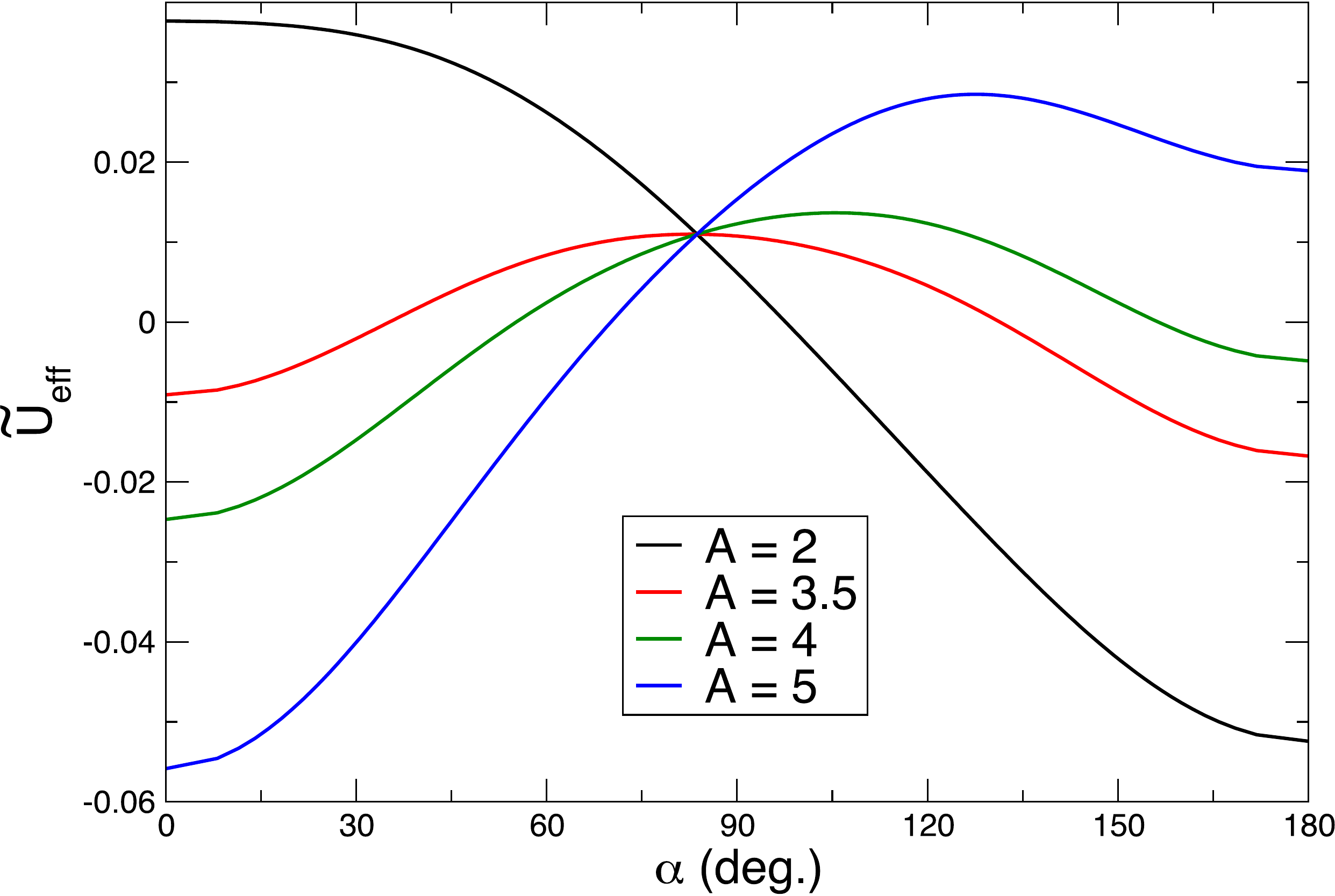}
\caption{\label{fig:eff_pots_withGrav} Effective potential $\tilde{U}_{eff}$ for a half-covered, phototactic ($\tilde{b} = 1.1$), bottom-heavy ($G = 0.1$) particle that is illuminated from above, for various values of the activity parameter $A$. The functional form of the effective potential (e.g., the location of the minimum) changes with $A$, due to the competition between bottom-heaviness and phototaxis.}
\end{figure}

Typical experimental realizations of light-activated Janus particles are bottom-heavy, e.g., when the particle consists of a catalytic metal film deposited on an inert spherical core. In the following, we assume that the direction of gravity is either $\hat{g} = -\hat{z}'$, i.e., the particle is illuminated from below, or $\hat{g} = \hat{z}'$, i.e., the particle is illuminated from above.  In either case, we can easily include the effect of bottom-heaviness on the distribution of orientations:
\begin{equation}
\label{eq:boltzmann_half_bh}
P(\alpha) = \mathcal{N} \exp \left[ -Pe^{r} \, (\tilde{U}_{eff}(\alpha) + \tilde{U}_{bh}(\alpha)) \right]
\end{equation}
where the dimensionless gravitational potential is 	
\begin{equation}
\label{eq:U_bh}
\tilde{U}_{bh}(\alpha) = (\hat{g} \cdot \hat{z}') \, \frac{R}{V_{s}} \frac{\tau_{bh}^{0}}{8 \pi \mu R^{3}} \cos(\alpha),
\end{equation}
and where $\tau_{bh}^{0}$ is the maximum value of the torque due to bottom-heaviness. We define another dimensionless parameter, ${G}$, as the prefactor of $\tilde{U}_{bh}$:
\begin{equation}
{G} \equiv \frac{R}{V^{s}} \frac{\tau_{bh}^{0}}{8 \pi \mu R^{3}}.
\end{equation}
One interesting aspect of the interplay of bottom-heaviness and activity is that the functional form of the ``potential landscape'' $U(\alpha) = U_{eff}(\alpha) + U_{bh}(\alpha)$ can be tuned by changing the illumination intensity, as parameterized by $A$, since $U_{bh}$ does not depend on illumination. Secondly, in certain cases, bottom-heaviness can compete with activity: consider, for instance, particles that are phototactic and illuminated from above. Bottom-heaviness tends to orient the particle orientation $\mathbf{\hat{d}}$ with the vertical; recalling that $\alpha$ is the angle of $\mathbf{\hat{d}}$  with the direction of light, bottom-heaviness drives $\alpha$ towards $\alpha = 180^{\circ}$.
Phototaxis rotates the vector $\mathbf{\hat{d}}$  to align with the direction of illumination, i.e., it drives $\alpha$ towards $\alpha = 0^{\circ}$. (Similar considerations hold for ``anti-phototactic'' particles ($\tilde{b} < 1$) when the particle is illuminated from below.)  In Fig. \ref{fig:PDFs_phototactic_lightAbove}, we show probability distributions of the particle orientation $\alpha$ for $\tilde{b} = 1.1$ and various values of $A$ when $G = 0.1$ at $\hat{g} = \hat{z}'$, i.e., the particle is illuminated from above.  In Fig. \ref{fig:eff_pots_withGrav}, we show the corresponding effective potential landscapes. Interestingly, the competition between phototaxis and bottom-heaviness can lead, at certain values of $A$, to bistability in the potential landscape $U_{eff}$, and hence to a bimodal distribution of orientations.

\begin{figure}[htb]
\includegraphics[width=\columnwidth]{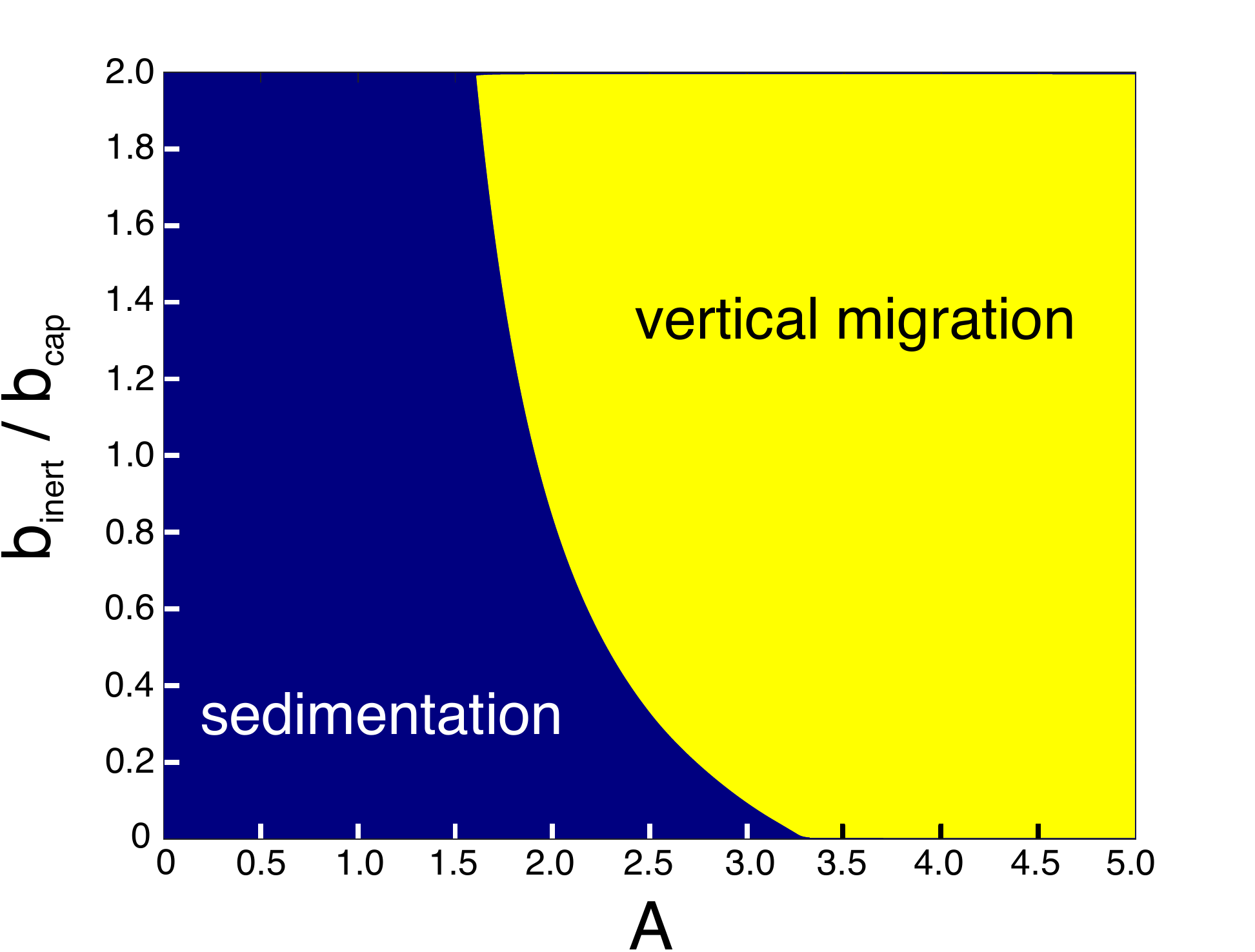}
\caption{\label{fig:phasemap_UzSwim_small} Phase map distinguishing sedimentation and vertical migration for the ``small'' particles ($G = 0.1$ and $Pe = 1$) as a function of $\tilde{b} > 0$ and $A$. The phase map is calculated for $b_{cap} < 0$. The particles are illuminated from below.}
\end{figure}

Now we consider the vertical migration of a photoactive particle, illuminated from below, against gravity.  Working in the stationary frame, defined in the discussion preceding Eq. (\ref{eq:Uxprime}), the particle has a net  velocity $\mathbf{U}_{tot} = \mathbf{U}_{swim}(\alpha) - V^{s} \hat{z}'$. The components of $\mathbf{U}_{swim}(\alpha)$ are given as $\mathbf{U}$ in Eqs. (\ref{eq:Uxprime})-(\ref{eq:Uzprime}). The time-averaged vertical component of the swimming velocity  can be calculated from the effective potential as:
\begin{equation}
\left< U^{z'}_{swim} \right>  = \int P(\alpha) \: U^{z'}_{swim}(\alpha) \: \sin \alpha \: d\alpha.
\end{equation}
In order for a particle to migrate vertically i.e., escape gravity,  the quantity $\left< U^{z'} \right>$ must satisfy $\left< U^{z'}_{swim} \right> = V^{s}$.  We consider two sets of parameters $G$ and $Pe$, corresponding to the ``small'' and ``large'' particles of Ref. \citenum{singh18}. The small particles have $G = 0.1$ and $Pe = 1$, while the large particles have $G = 0.05$ and $Pe = 20$.  The phase map in Fig. \ref{fig:phasemap_UzSwim_small} shows whether the small particles sink or migrate vertically as a function of the parameters $A$ and $\tilde{b}$ for $\tilde{b} > 0$ and $b_{cap} < 0$. Regardless of whether the particles are phototactic ($\tilde{b} > 1$) or anti-phototactic  ($\tilde{b} < 1$), the particles migrate vertically for $A$ larger than a threshold value.  

\begin{figure}[htb]
\includegraphics[width=\columnwidth]{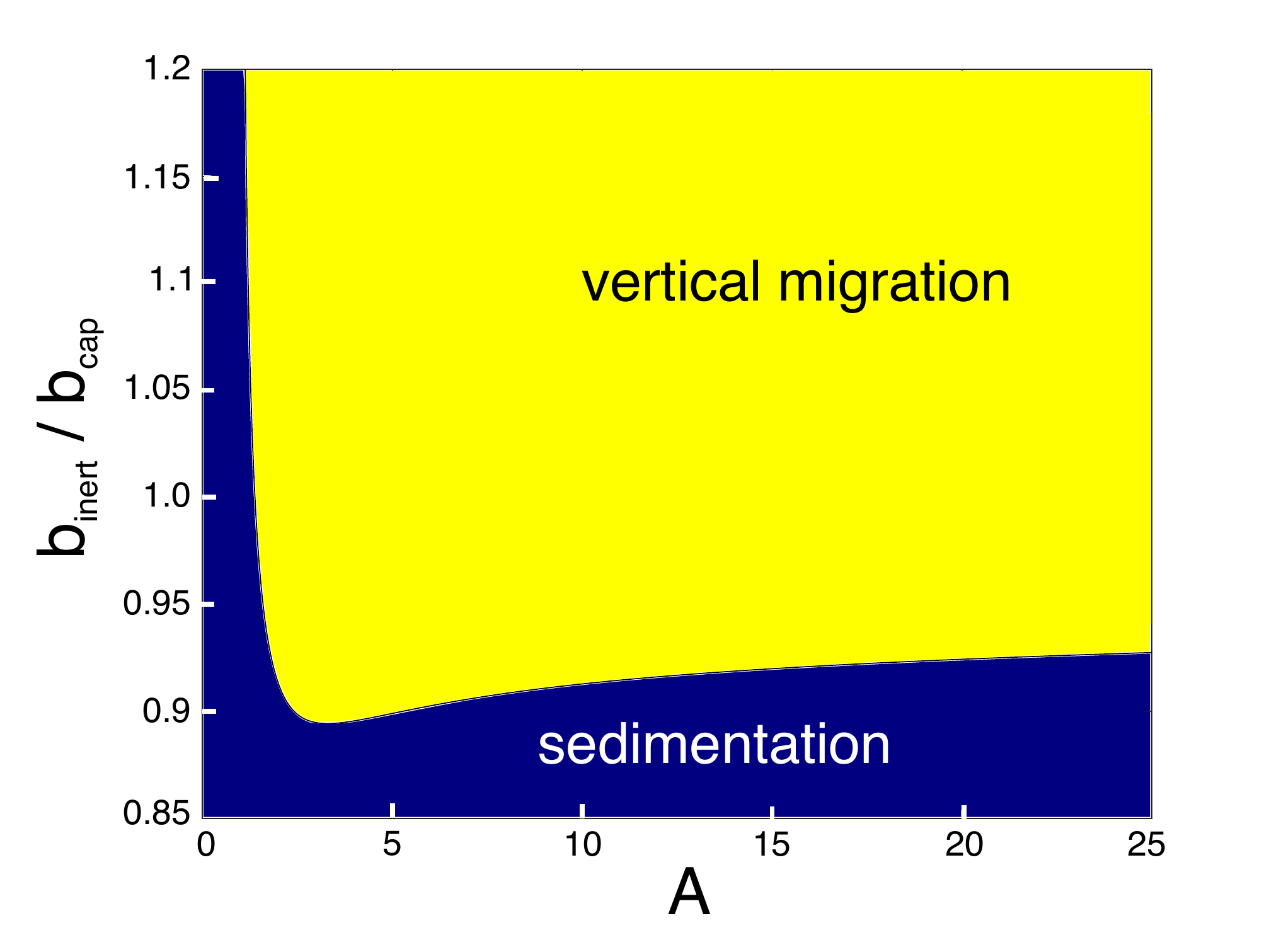}
\caption{\label{fig:phasemap_UzSwim_large} Phase map distinguishing sedimentation and vertical migration for the ``large'' particles ($G = 0.05$ and $Pe = 20$) as a function of $\tilde{b} > 0$ and $A$. The phase map is calculated for $b_{cap} < 0$.  The particles are illuminated from below.}
\end{figure}

Fig. \ref{fig:phasemap_UzSwim_large} shows a phase map for the large particles. The behavior for phototactic ($\tilde{b} > 1$) particles is similar as for small particles. However, for anti-phototactic particles  ($\tilde{b} < 1$), there is a range of $\tilde{b}$ for which the phase behavior is re-entrant: with increasing $A$, the behavior passes from sedimentation to vertical migration and back to sedimentation. The reason for this is the following: at low $A$, the orientation of these particles  is dominated by bottom-heaviness, such that the average orientation of the particle is vertical. In this region, increasing $A$ increases the mean vertical swimming velocity. However, for larger values of $A$, the anti-phototactic effect becomes as significant as bottom-heaviness, and the average orientation of the particles shifts away from upward alignment towards downward alignment.

\begin{figure}[htb]
\includegraphics[width=\columnwidth]{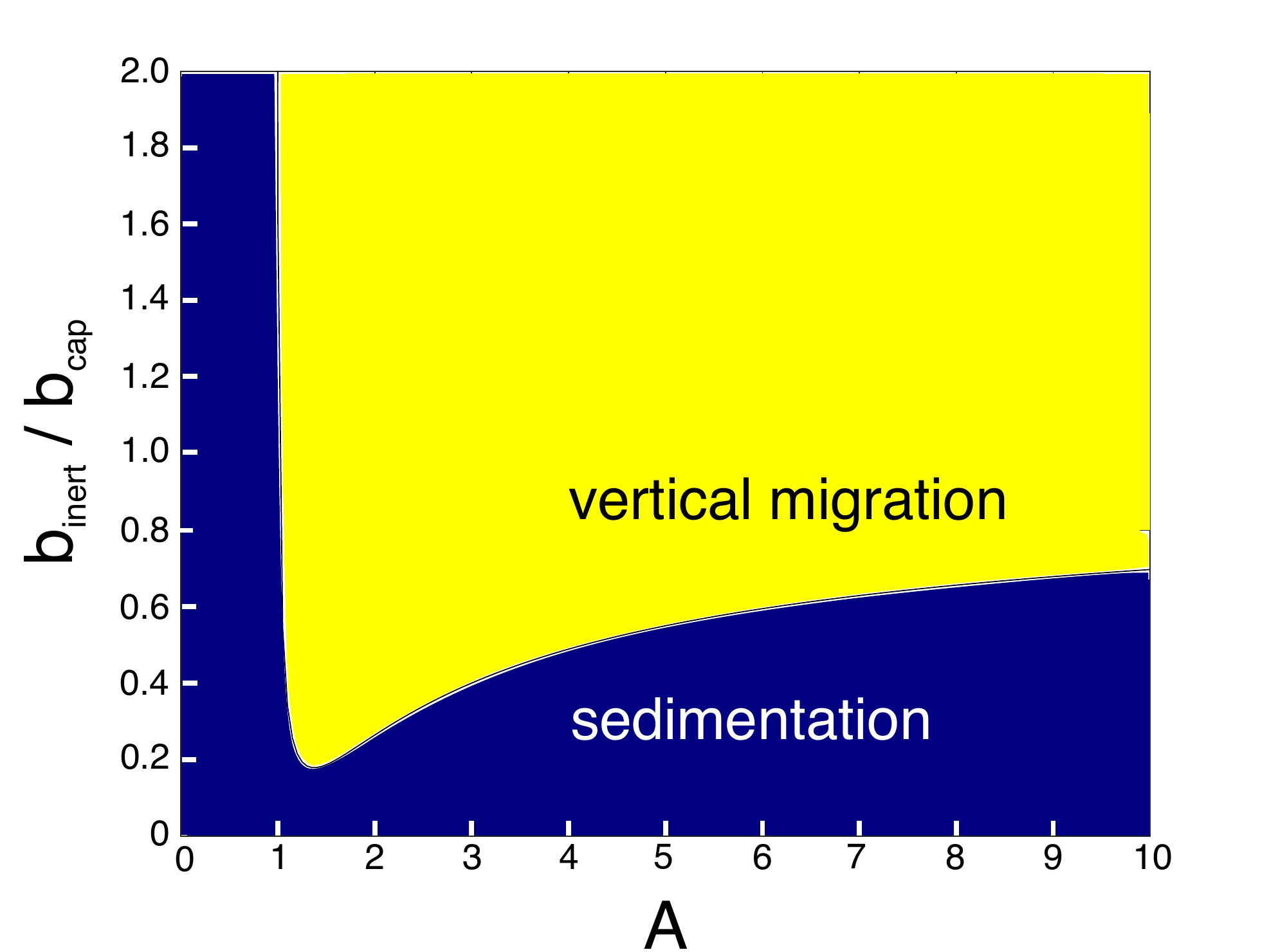}
\caption{\label{fig:phasemap_UzSwim_large2} Phase map distinguishing sedimentation and vertical migration for particles  with $G = 0.5$ and $Pe = 20$ as a function of $\tilde{b} > 0$ and $A$. The phase map is calculated for $b_{cap} < 0$. The particles are illuminated from below.}
\end{figure}

The re-entrant region is small. In order to demonstrate that it can be larger, we consider a particle with $Pe = 20$ -- same as the large particles -- but with $G = 0.5$. In Fig. \ref{fig:phasemap_UzSwim_large2}, the re-entrant region of the phase map extends over a wide range of $\tilde{b}$. 

Finally, we consider the observable $\left< U^{z'}_{swim} \right>$ in more quantitative detail.  In addition, we also define
\begin{equation}
\left< U^{\perp}_{swim} \right>  = \int P(\alpha) \: U^{\perp}_{swim}(\alpha) \: \sin \alpha \: d\alpha,
\end{equation}
where
\begin{equation}
U^{\perp}_{swim}(\alpha) \equiv U^{d} \sin(\alpha) + U^{p} \cos(\alpha). 
\end{equation}
The quantity $U^{\perp}_{swim}$ represents the  \textit{magnitude} of the  component of the swimming velocity in the plane normal to $\mathbf{\hat{q}}$ .  (By symmetry,  $ \left< U^{x'} \right> = 0$ and $ \left< U^{y'} \right> = 0$.) 

\begin{figure}[htb]
\includegraphics[width=\columnwidth]{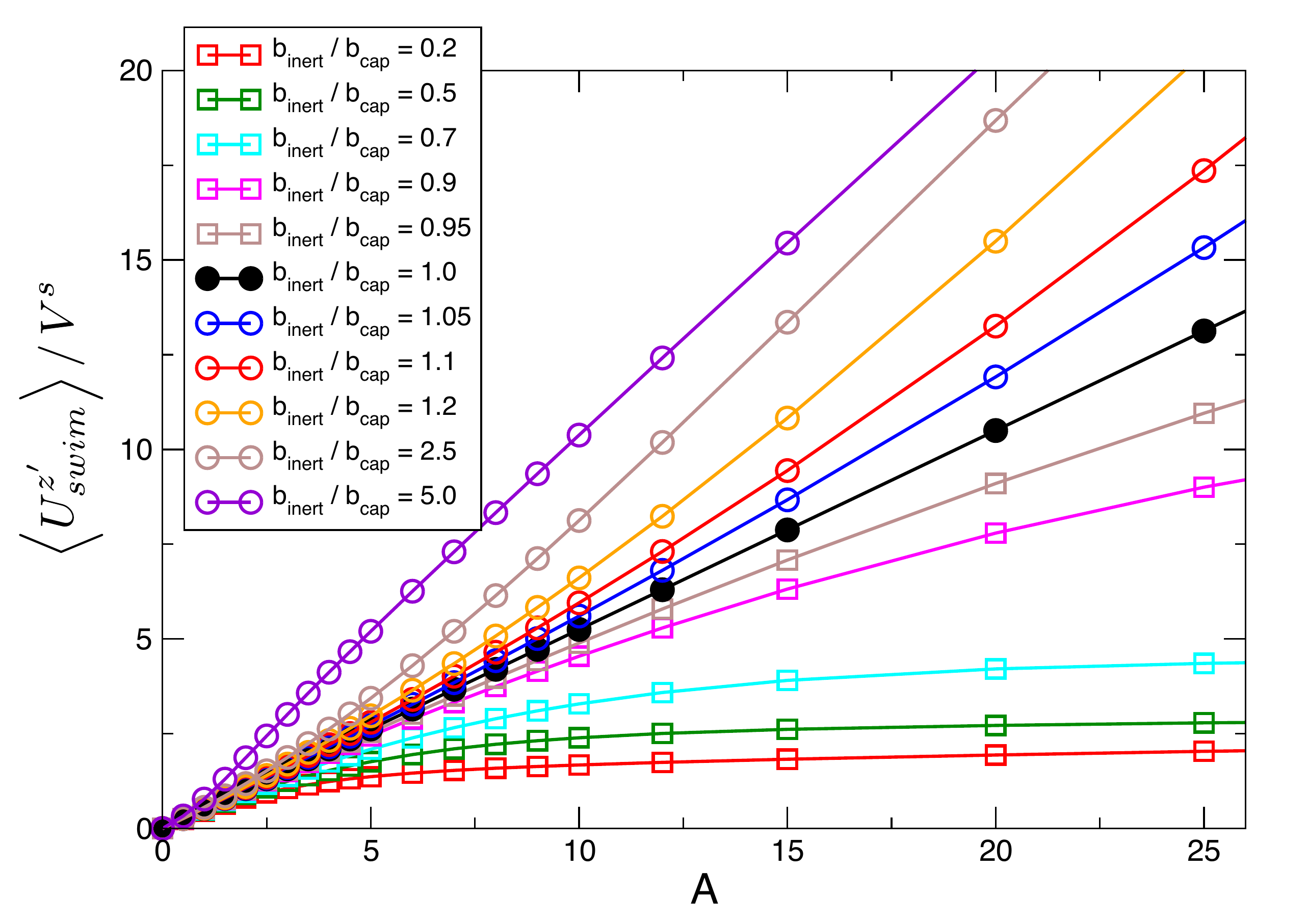}
\caption{\label{fig:UzSwim_small} Mean vertical component of the swimming velocity $\left< U^{z'}_{swim} \right>$ as a function of $A$ for the  small particles ($G = 0.1$ and $Pe = 1$), at various values of $\tilde{b}$. The particles are illuminated from below.  }
\end{figure}

In Fig. \ref{fig:UzSwim_small}, we show $\left< U^{z'}_{swim} \right>$ as a function of $A$ for various values of $\tilde{b}$. For $\tilde{b} = 1$, there is no phototactic effect, and this function is linear. For phototactic particles ($\tilde{b} > 1$), the function is weakly superlinear, while for anti-phototactic particles ($\tilde{b} < 1$),  it is weakly sublinear for low $A$. The nonlinearity of $\left< U^{z'}_{swim} \right>$ for phototactic and anti-phototactic  particles reflects the fact that photo-alignment is somewhat weak for low values of $A$ (i.e., less significant than, or comparable, to the effects of fluctuations and bottom-heaviness), and hence the mean orientation can shift significantly as $A$ changes. Furthermore, the mean vertical swimming velocity is non-monotonic for the anti-phototactic particles: the velocity peaks and then decreases with increasing $A$, reflecting the dominance of anti-phototaxis over bottom-heaviness at high $A$. This non-monotonic form is responsible for the re-entrant phase behavior discussed above. 

\begin{figure}[htb]
\includegraphics[width=\columnwidth]{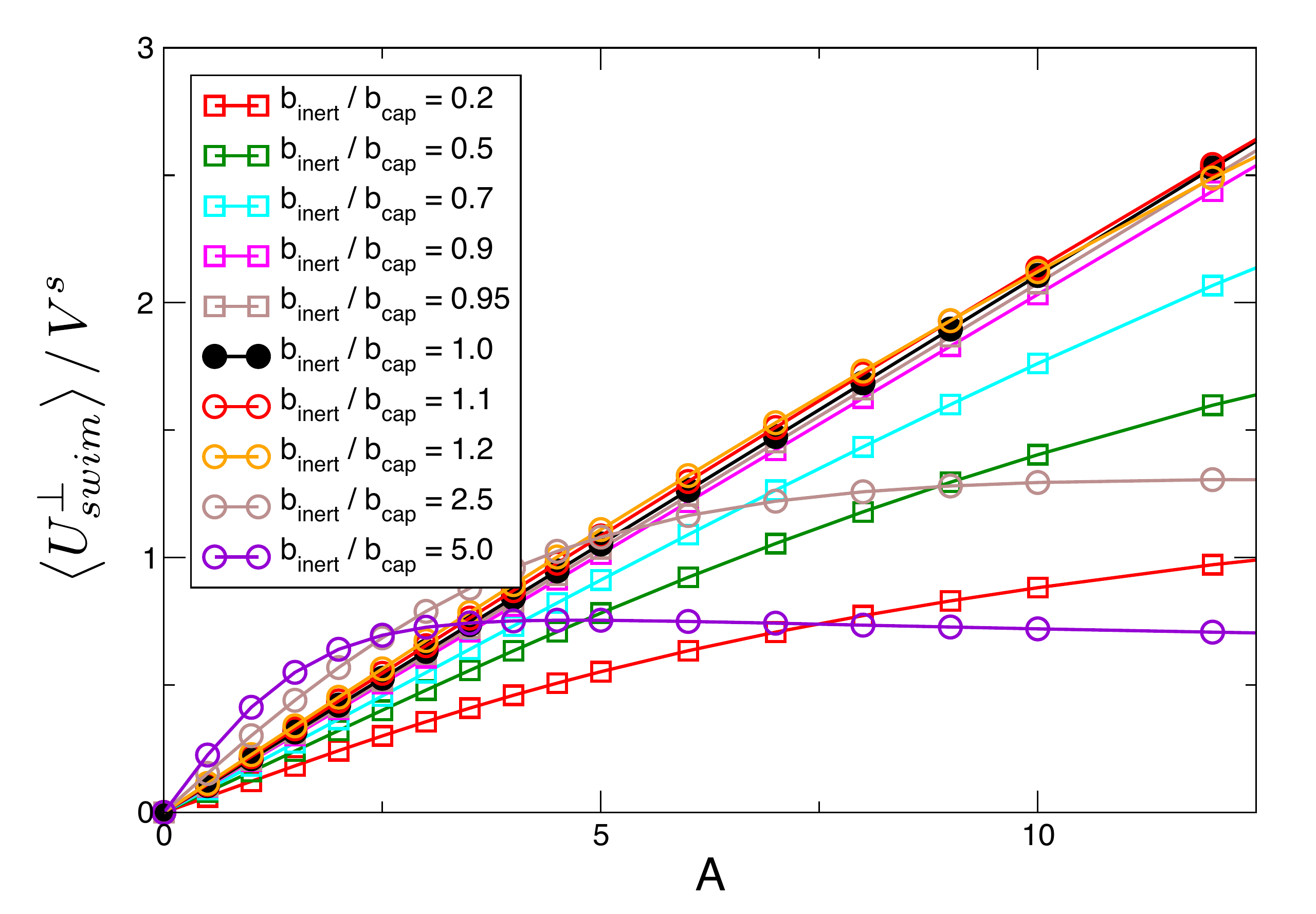}
\caption{\label{fig:UxySwim_small}   Mean perpendicular component of the swimming velocity $\left< U^{\perp}_{swim} \right>$ as a function of $A$ for the small particles ($G = 0.1$ and $Pe = 1$), at various values of $\tilde{b}$. The particles are illuminated from below.  }
\end{figure}

In Fig. \ref{fig:UxySwim_small}, we show the mean perpendicular component of particle velocity as a function of activity parameter $A$ for the small particles. For both phototactic and anti-phototactic particles, this function is non-monotonic: it initially increases with $A$, reaches a peak, and thereafter decreases with $A$. The initial increase with $A$, which also holds for the non-tactic particles ($\tilde{b} = 1$), is due to the general increase of swimming speed as $A$ is increased. At high $A$, phototactic or anti-phototactic alignment becomes dominant,  and as $A$ is increased, tactic alignment in $ \pm \mathbf{\hat{q}}$ is strengthened, decreasing $\left< U^{\perp}_{swim} \right>$.

\begin{figure}[htb]
\includegraphics[width=\columnwidth]{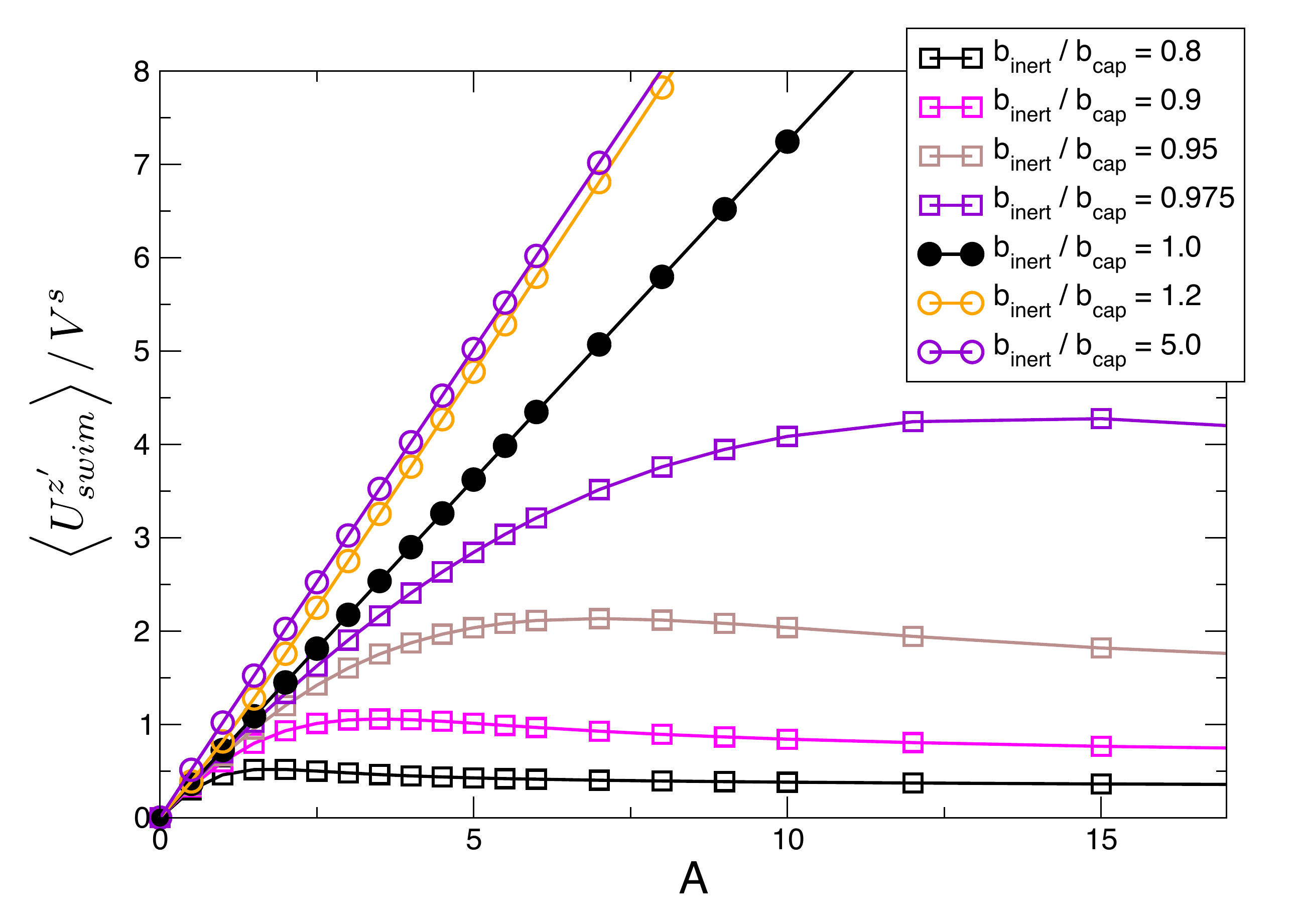}
\caption{\label{fig:UzSwim_large}   Mean vertical component of the swimming velocity $\left< U^{z'}_{swim} \right>$ as a function of $A$ for various values of $\tilde{b}$ for the large particles ($G = 0.05$ and $Pe = 20$). The particles are illuminated from below.  }
\end{figure}

\begin{figure}[htb]
\includegraphics[width=\columnwidth]{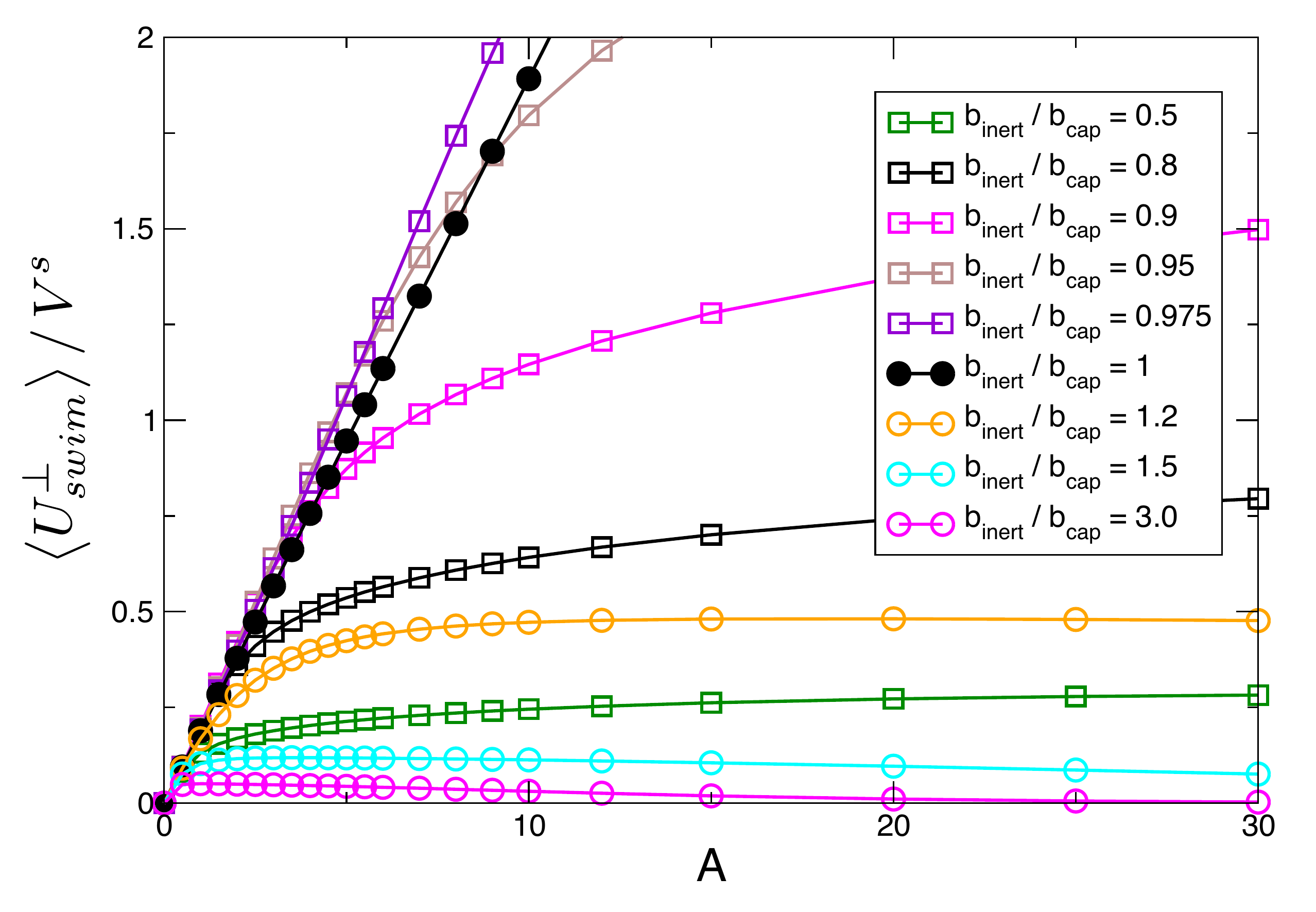}
\caption{\label{fig:UxySwim_large}   Mean perpendicular component of the swimming velocity $\left< U^{\perp}_{swim} \right>$ as a function of $A$ for various values of $\tilde{b}$ for the large particles ($G = 0.05$ and $Pe = 20$). The particles are illuminated from below.  }
\end{figure}

Now we consider the mean vertical velocity of the large particles in Fig. \ref{fig:UzSwim_large}. For large phototactic particles, alignment with the vertical is very strong for even low values of $A$, and the function is approximately linear. For large anti-phototactic particles, the nonlinearity of $\left< U^{z'}_{swim} \right>$ as a function of $A$ is more pronounced, and the peaks are clearly visible. We also show $\left< U^{\perp}_{swim} \right>$  for the large particles in Fig. \ref{fig:UxySwim_large}.

\subsection{Effective potential for a low coverage particle}

\begin{figure}[htb]
\includegraphics[width=0.9\columnwidth]{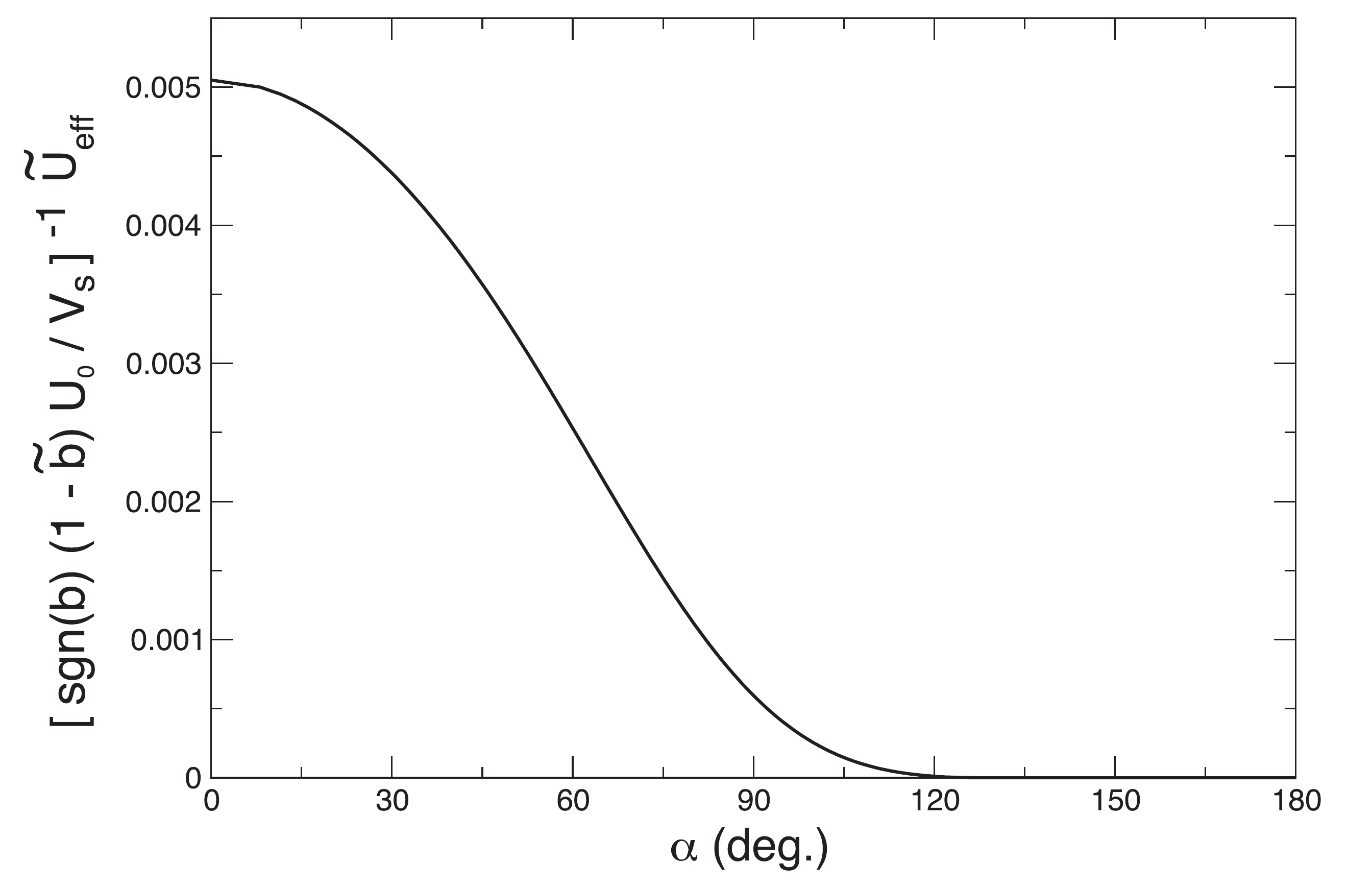}
\caption{\label{fig:chi0_n08_Ueff} Rescaled effective potential $\tilde{U}_{eff}$ for a particle with $\chi_0 = -0.8$.}
\end{figure}

The low coverage particles are interesting in the context of thermal fluctuations, due to the broad range of angles $\alpha \in [\alpha_{c2}, \pi]$ in which \textcolor{black}{$\Omega_y = 0$}. One can anticipate that $P(\alpha) = \textrm{const}$ in this region, and that $P(\alpha)$ is some function of $\alpha$ for $0 \leq \alpha < \alpha_{c2}$. We define a dimensionless effective potential for the low coverage particles as the following:
\begin{multline}
\label{eq:U_eff_low}
\tilde{U}_{eff}^{low}(\alpha) = \sgn(b_{cap}) (1 - \tilde{b}) \,  \Theta(\alpha_{ac2} - \alpha) \, \frac{U_0}{V^s} \\  \sum_{n \geq 1} \frac{a_{n} \alpha_{c2}}{n \pi} \left[ \cos(n \, \alpha \, \pi / \alpha_{c2}) - (-1)^n \right]
\end{multline}
where $\Theta(\alpha_{ac2} - \alpha_{ac2}^{-}) = 1$.
The effective potential master curve for $\chi_0 = -0.8$ and $G = 0$, obtained with the numerically fitted coefficents from Table I, is shown in Fig. \ref{fig:chi0_n08_Ueff}.  In Fig. \ref{fig:PDFs_phototactic_lowCov}, we show the probability distributions for the orientation of a phototactic particle (here with $\tilde{b} = 2.0$) with $\chi_0 = -0.8$ for various values of $Pe^{r} A$. The constant region of the effective potential indeed leads to a flat region of the probability distribution function.

\subsection{Brownian dynamics}

\begin{figure}[htb]
\includegraphics[width=\columnwidth]{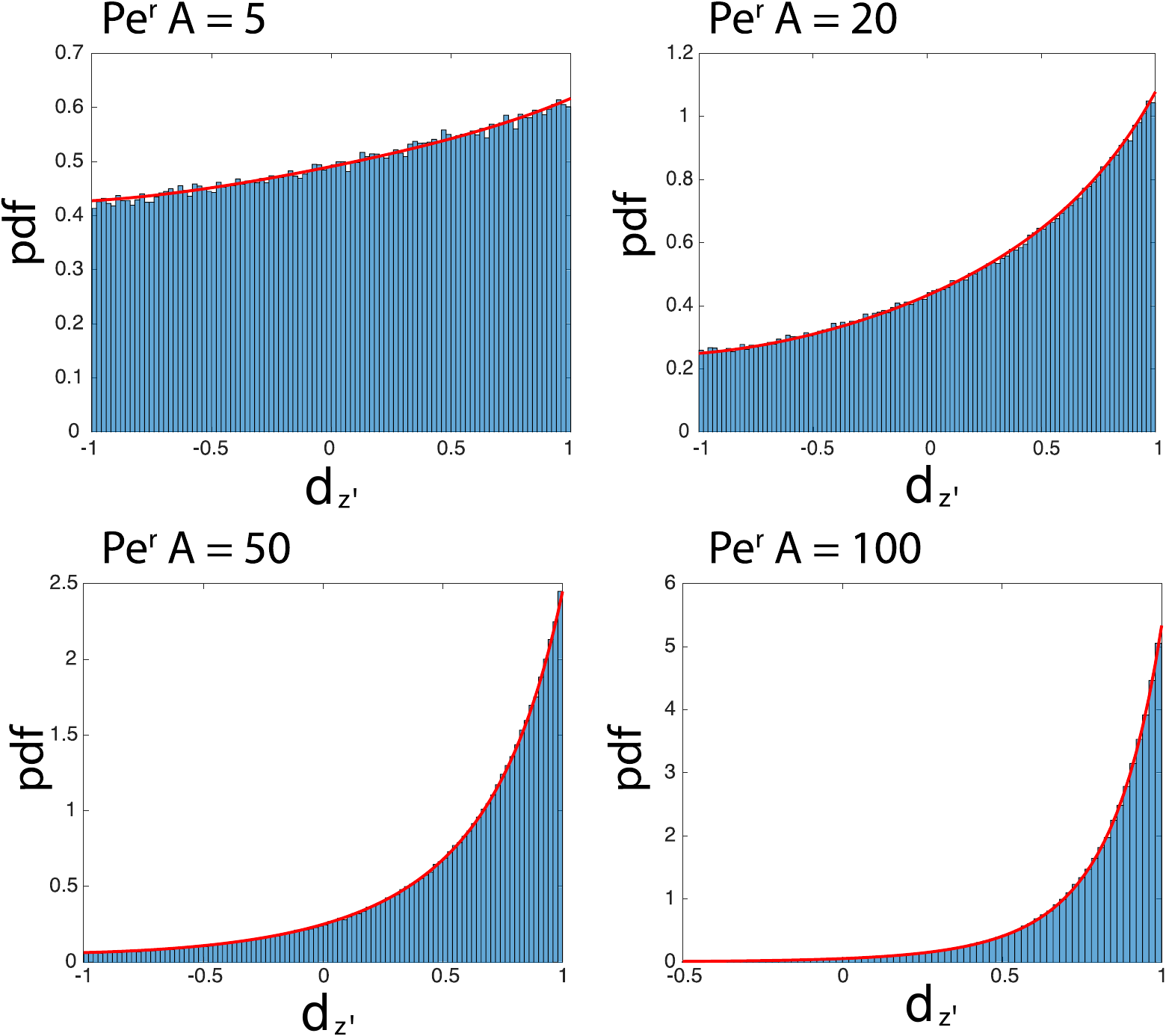}
\caption{\label{fig:PDFs_phototactic_halfCov} Probability distributions for the vertical component of orientation $d_{z'} = \cos \alpha$ for a half-covered, phototactic ($\tilde{b} = 1.1$) Janus particle at various values of $Pe^{r} A$. The red solid line shows the theoretical probability distribution, computed from Eq. (\ref{eq:boltzmann_half}) and Eq. (\ref{eq:tilde_U_eff_2}). The blue histogram was obtained from Brownian dynamics simulations. Bottom-heaviness is omitted ($G = 0$).}
\end{figure}

\begin{figure}[htb]
\includegraphics[width=\columnwidth]{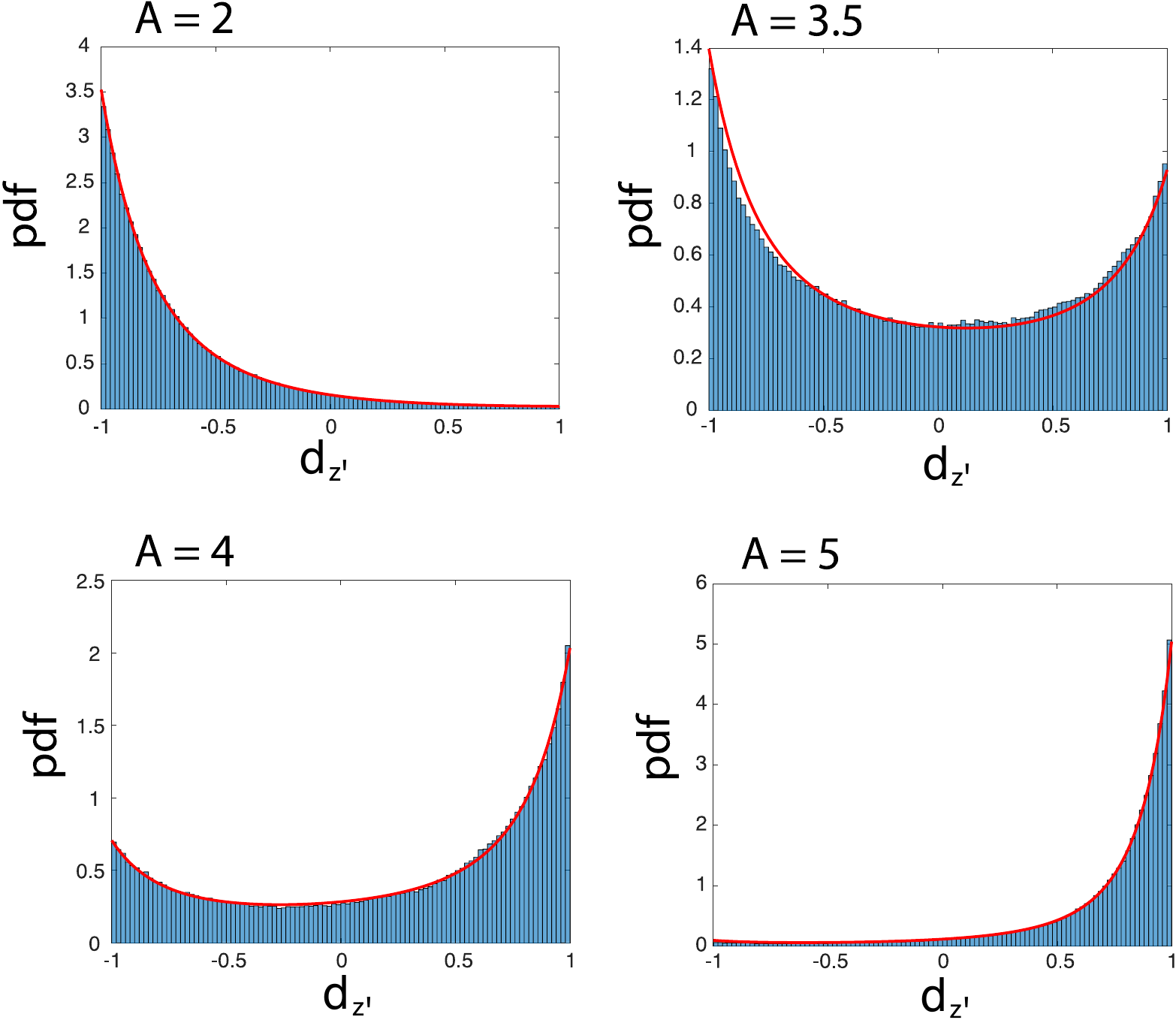}
\caption{\label{fig:PDFs_phototactic_lightAbove} Probability distributions for the vertical component of orientation $d_{z'} = \cos \alpha$ for a half-covered, phototactic ($\tilde{b} = 1.1$) Janus particle for various values of $A$. The rotational P\'{e}clet number is $Pe^{r} = 40$, and the particle is bottom-heavy ($G = 0.1$). The particle is illuminated from above ($\hat{g} \cdot \hat{z}' = 1$). We recall that $d_{z'} = 1$ corresponds to an alignment of the particle axis with the direction of light, so in this case, $d_{z'} = -1$ corresponds to alignment with the vertical direction (as defined by gravity). The red solid line shows the theoretical probability distribution, computed from Eq. (\ref{eq:tilde_U_eff}), Eq. (\ref{eq:boltzmann_half_bh}), and Eq. (\ref{eq:U_bh}). The blue histogram was obtained from Brownian dynamics simulations. }
\end{figure}

\begin{figure}[htb]
\includegraphics[width=\columnwidth]{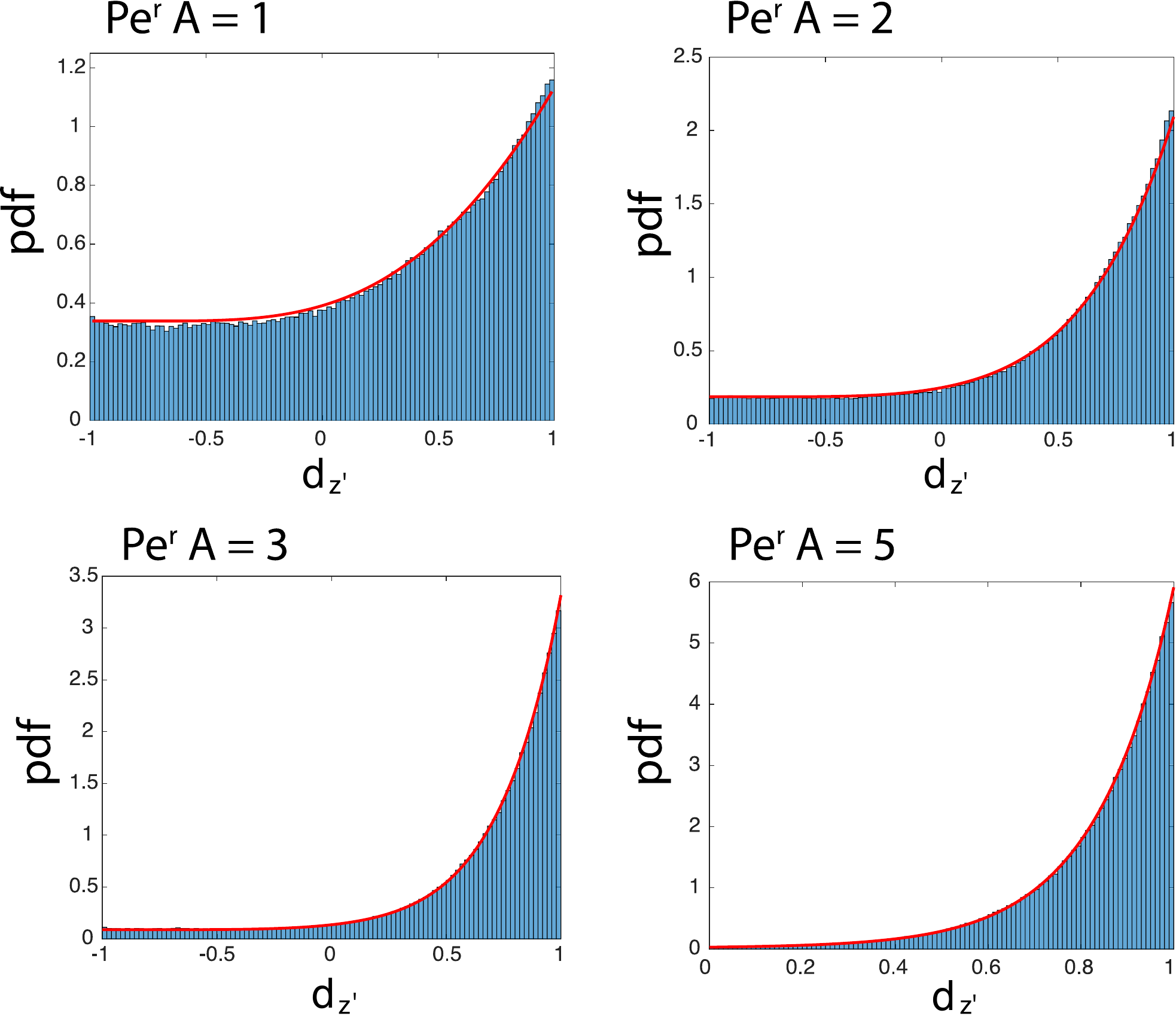}
\caption{\label{fig:PDFs_phototactic_lowCov} Probability distributions for the vertical component of orientation $d_{z'} = \cos \alpha$ for a phototactic ($\tilde{b} = 2.0$) Janus particle with low catalyst coverage ($\chi_0 = -0.8$) at various values of $Pe^{r} A$. The red solid line shows the theoretical probability distribution, computed from Eq. (\ref{eq:boltzmann_half}) and Eq. (\ref{eq:U_eff_low}). The blue histogram was obtained from Brownian dynamics simulations. Bottom-heaviness is omitted ($G = 0$). Because the catalytic cap is completely in shadow for the angles $\alpha_{ac2} < \alpha < \pi$, corresponding to $-1 < d_{z'} < -0.6$, there are nearly flat regions of the probability distribution for low values of $Pe^{r} A$ ($Pe^{r} A = 1$, $Pe^{r} A = 2$, and $Pe^{r} A = 3$.)}
\end{figure}

The Brownian dynamics simulations are performed as follows. The particle has a time-dependent position $\mathbf{r}_{p}(\tilde{t})$ and a time-dependent orientation $\mathbf{\hat{d}}(\tilde{t})$, where $|\mathbf{\hat{d}}| = 1$, the dimensionless time $\tilde{t}$ is $\tilde{t} \equiv t/T_0$, and the characteristic timescale $T_{0}$ is $T_{0} \equiv R/U_{0}$. At the beginning of a simulation, we specify the size of the (dimensionless) simulation timestep, $\Delta \tilde{t} \equiv \Delta t/T_{0}$. We always take $\Delta \tilde{t} \leq 0.01$. We specify the initial position as $\mathbf{r}_{p}(0) = (0, 0, 0)$, and the initial orientation is selected at random. The calculations are peformed in the stationary, ``primed'' frame in which the $\hat{z}'$ axis is aligned with or against the direction of light, i.e., $\hat{z}' = \pm \mathbf{\hat{q}}$ as discussed above.

We run each simulation for $N$ timesteps, choosing $N$ such that $N \Delta \tilde{t} \geq 50000$. At each timestep, we define $\mathbf{\hat{d}}_{0} = \mathbf{\hat{d}}(t)$.  We compute $\mathbf{U}(\mathbf{\hat{d}}_{0})$ and $ \bm{\Omega}(\mathbf{\hat{d}}_0)$ from either analytical expressions (for the half-covered particle), or by interpolating from numerical data obtained with the BEM (for other coverage levels). We obtain the position at the next timestep as
\begin{equation}
\mathbf{r}_{p}(\tilde{t} + \Delta \tilde{t}) = \mathbf{r}_{p}(\tilde{t}) + \left[ \mathbf{U}_{swim}(\mathbf{\hat{d}}_{0}) + \mathbf{V}_{s} \right] \Delta \tilde{t} + \mathbf{s}_{t}.
\end{equation}
Here, $\mathbf{s}_{t}$ is a vector of three Gaussian distributed independent random variables with $\left< s_{t,i} \right> = 0$ and $\left< s_{t,i} s_{t,j} \right> = 2 Pe_{p}^{-1} \Delta \tilde{t} \delta_{ij}$. The quantity $\mathbf{U}^{swim}$ is the translational velocity due to activity, and $\mathbf{V}_{s}$ is the sedimentation velocity. Concerning the orientation, we first compute the deterministic change in the orientation vector $\dot{\mathbf{\hat{d}}}$ from 
\begin{equation}
\mathbf{\dot{\hat{d}}} = \left[ \bm{\Omega}_{swim}(\mathbf{\hat{d}}_0) + \bm{\Omega}_{bh}(\mathbf{\hat{d}}_0) \right] \times \mathbf{\hat{d}}_0.
\end{equation}
Here, $\bm{\Omega}_{swim}$ is the angular velocity due to activity, and $\bm{\Omega}_{bh}$ is the angular velocity due to bottom-heaviness (if included.) We then compute
\begin{equation}
\mathbf{{d}}' = \mathbf{\hat{d}}_{0} + \mathbf{\dot{\hat{d}}} \Delta \tilde{t}   + \mathbf{s}^{r} \times \mathbf{\hat{d}}_{0}.
\end{equation}
Here, $\mathbf{s}_{r}$ is a vector of three Gaussian distributed independent random variables with $\left< s_{r,i} \right> = 0$ and $\left< s_{r,i} s_{r,j} \right> = 2 Pe_{r}^{-1} \Delta \tilde{t} \delta_{ij}$.  After obtaining  $\mathbf{{d}}'$, we normalize the length of $\mathbf{{d}}'$ to obtain the orientation vector at the next time step:
\begin{equation}
\mathbf{\hat{d}}(\tilde{t} + \Delta \tilde{t}) = \frac{\mathbf{{d}}'}{|\mathbf{{d}}'|}.
\end{equation}
The probability distributions for the particle orientation are obtained from $\mathbf{\hat{d}}(\tilde{t})$ for $\tilde{t} > 100$. These are shown as blue histograms in Figs. \ref{fig:PDFs_phototactic_halfCov}, \ref{fig:PDFs_phototactic_lightAbove}, and \ref{fig:PDFs_phototactic_lowCov}, and show excellent agreement with the theoretically predicted Boltzmann distributions.


\begin{figure}[htb]
\includegraphics[width=1.0\columnwidth]{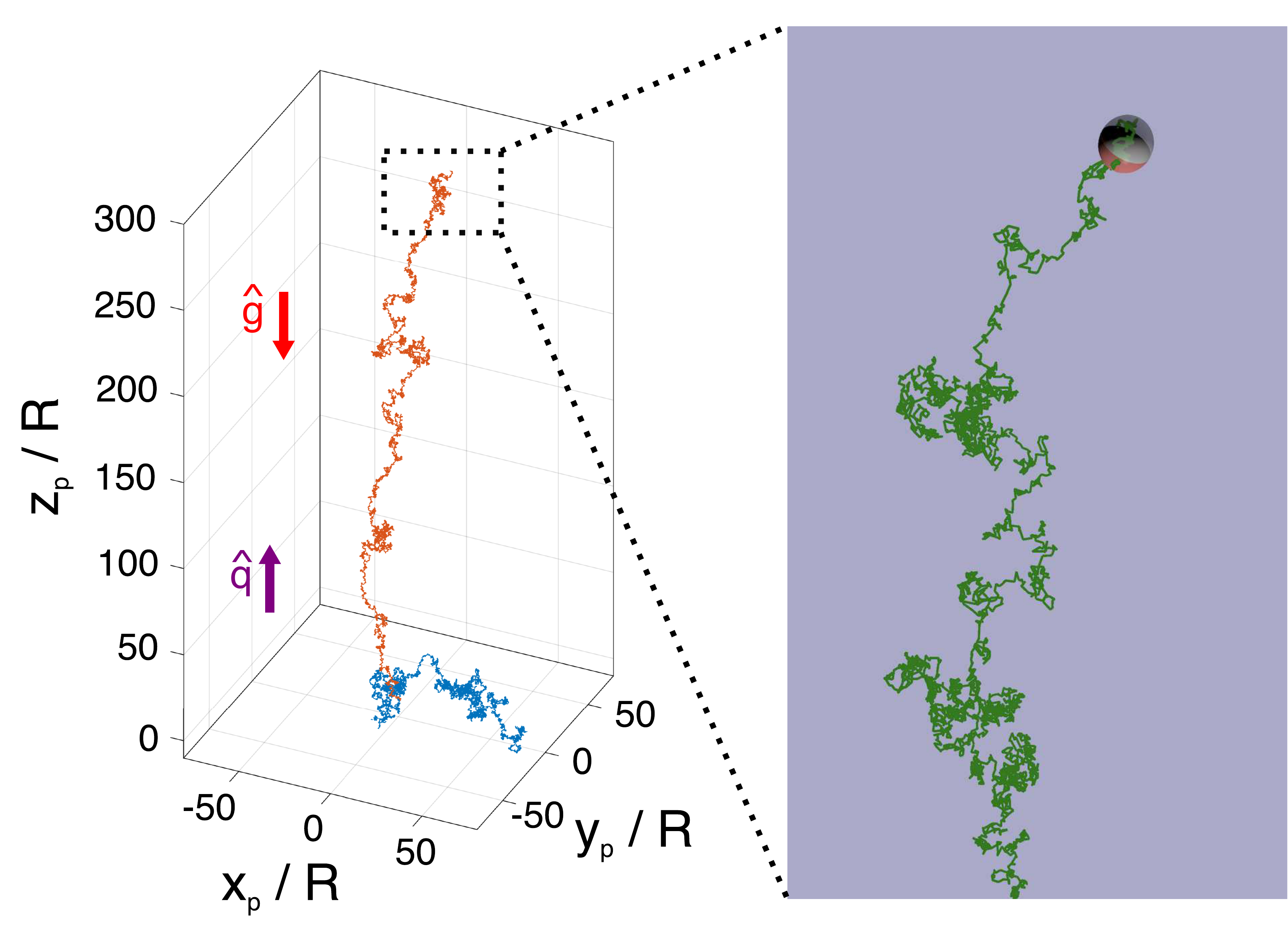}
\caption{\label{fig:3dtraj} (left) Three-dimensional trajectories of a particle with $\tilde{b} = 1.1$ and $G = 0.05$, corresponding to the ``small'' particles of Ref.\citenum{singh18},  for $A = 1.9$ (blue) and $A = 3$ (orange). The particle is illuminated from below. For $A = 1.9$, the average  vertical velocity of the particle is zero, and the particle exhibits a quasi-2D random walk near the $z = 0$ plane. For $A = 3$, the bottom-heavy and phototactic particle migrates against the direction of gravity. (right) A zoomed-in view of the trajectory  with $A = 3$.}
\end{figure}


Finally, we note that our approach can be used to obtain particle trajectories. For instance, in Fig. \ref{fig:3dtraj}, we show characteristic trajectories for a particle with $\tilde{b} = 1.1$, $G = 0.05$, corresponding to the ``small'' particles of Ref.\citenum{singh18}, illuminated from below with $A = 1.9$  (blue) and $A = 3$ (orange). The phototactic particles are illuminated from below. For $A = 1.9$, the average vertical swimming velocity $\left<U_{swim}^{z'}\right> \approx V_s$, making $U_{tot}^{z'} \approx 0$. (See Fig. \ref{fig:phasemap_UzSwim_small}.) However, $\left<U_{swim}^{\perp}\right> \neq 0$, and therefore the particle exhibits a quasi-2D random walk largely confined to the $xy$ plane. For $A = 3$,  the particle swims fast enough to overcome the sedimentation velocity, and it migrates vertically. 

\section{Conclusions}

In this manuscript, we developed a theory of the motion of light-activated catalytic Janus particles in bulk solution.  We considered a particle with a catalytic cap that is opaque to light, corresponding to some recent experimental studies, which leads to an orientation-dependent ``self-shadowing'' effect. Working within a continuum framework for self-diffusiophoretic motion, we obtained analytical expressions for the deterministic contributions to  translation and rotation of a half-covered particle. In particular, a particle can rotate its catalytic cap towards (phototaxis) or away from (anti-phototaxis) the incident light, depending on its surface chemistry. Our analytical results were confirmed by detailed numerical calculations. Numerically, we investigated the effect of changing the level of catalyst coverage, and found that the extent of coverage can have a significant effect on the functional form of the rotational velocity as a function of orientation. For instance, particles with low coverage have a broad range of orientations in which the cap is completely in shadow and the rotational velocity is zero.

We then considered the effect of thermal fluctuations. We found that the orientation of the particle obeys a Boltzmann distribution with a nonequilibrium effective potential. When photoactivity is combined with bottom-heaviness, the overall effective potential changes its functional form as the light intensity is varied. These findings expose a novel route to light-tunable and light-programmable active particle motion. We also investigated the dynamical phase behavior of the photoactive particles, i.e., the conditions under which a particle, illuminated from below, will either sediment or migrate against gravity. We found that certain types of particle exhibit re-entrant dynamical phase behavior: for low light intensity, they sediment; for intermediate intensity, they migrate vertically; and for high intensity, they sediment. This prediction provides a clear signature of anti-phototaxis for experimental studies.

\textcolor{black}{It should be noted that, concerning the deterministic contributions to particle rotation, the only stable particle orientations observed in this manuscript were either with or against the direction of illumination. In the absence of bottom-heaviness or other contributions to rotation, this fact follows from the axial symmetry of the surface chemistry of the particle. When we considered the effect of bottom-heaviness, we assumed that the direction of incident light was aligned either with or against the vertical, which is typical of most experimental setups. Consequently, the only stable orientations were again either ``cap up'' or ``cap down,'' even though bottom-heaviness can (for some parameters) compete with tactic rotation. If the direction of illumination and the direction of gravity $\mathbf{g}$ were to have some angle $\gamma$ between them, with $0^{\circ} < \gamma < 180^{\circ}$,  it is conceivable that the particle would align in an intermediate direction, i.e., a direction between $\mathbf{\hat{q}}$ and $\mathbf{g}$ in the plane containing $\mathbf{\hat{q}}$ and $\mathbf{g}$.  A follow-up paper could perform a more general analysis for arbitrary  $\mathbf{\hat{q}}$ and $\mathbf{g}$. }

\textcolor{black}{Moreover,} this manuscript concerned the behavior of a single particle. For a dilute suspension of particles,  single-particle behavior will govern the spatial migration of the particle population. In a semi-dilute suspension, interactions between two or more particles could lead to rich new physics. First, interaction of the particles through their self-generated hydrodynamic and chemical fields will lead to new contributions to the translational and rotational velocities of a particle. These contributions could have a  complex interplay with the orientation dependence, studied in this manuscript, of the activity and velocity of a single particle. Secondly, if the vector between the centers of two particles is aligned with the direction of light propagation $\mathbf{\hat{q}}$, one particle will block the propagation of light to the other particle.  Recent studies have shown that this particle/particle shadowing can lead to interesting collective effects, including the emergence of comet-like swarms of isotropic photo-active particles,\cite{cohen14} and spatial focusing of a suspension of phototactic micro-organisms exposed to flow\cite{clarke18}. 

Throughout this manuscript we assumed that the particle is sufficiently large,  compared with the wavelength of incident light, that  geometrical optics provides an adequate description of the \textcolor{black}{illumination of the catalytic cap}. As the particle size is reduced to the submicron or nano regime, Mie and Rayleigh scattering effects will become important. These effects could be modeled analytically\textcolor{black}{\cite{bohren83}} or numerically\textcolor{black}{\cite{reid15}} (with the boundary element method), and could enrich single particle behavior or interactions between particles, especially if the scattered electromagnetic field is highly anisotropic due to the Janus character of the particle. 

\textcolor{black}{In order to assess quantitatively the validity of the geometrical optics approximation, we consider the ``size parameter'' $x \equiv 2 \pi R/\lambda$, which is commonly used in the Mie scattering literature. Here, $R$ is the radius of the particle and $\lambda$ is the wavelength of incident light. The geometrical approximation is considered valid in the range $x \gg 1$. An examination of the literature can give a more precise sense of when $x$ is sufficiently larger than one. For instance, the recent work of Masoumeh-Mousavi \textit{et al}. considers self-photothermophoretic Janus particles in a  Gaussian optical field.\cite{masoumeh18} They use particles with diameters of $D = 4.77 \; \mathrm{\mu m}$ and $D=6.73 \; \mathrm{\mu m}$ exposed to laser light with a wavelength of $\lambda = 976 \; \mathrm{nm}$, giving $x = 15.3$ and $x = 21.7$ for the two particle sizes. Notably, they successfully model the optical forces and torques on the particle using the geometrical optics approximation. Similarly, Park and Furst consider the optical forces on a microsphere in an optical trap.\cite{park14} They find that the geometrical optics approximation is accurate for dielectric microspheres with radii between $R=1.35\;\mathrm{\mu m}$ and  $R = 3.7\;\mathrm{\mu m}$ in light with wavelength $\lambda = 1064\; \mathrm{nm}$, i.e., for $x$ between $x = 7.97$ and $x = 21.8$. It should be noted that both studies consider the validity of the geometrical optics approximation for calculation of optical forces on a particle in an optical trap, rather than calculation of photocatalytic activity. There are fewer studies that address the validity of the approximation in the latter context. One example is from Gerischer and Heller, who consider TiO\textsubscript{2} semiconducting microspheres in water and exposed to sunlight.\cite{gerischer92} They consider the geometrical optics approximation to be appropriate for radii $R \ge 1 \; {\mu m}$ and  light with wavelength $\lambda=400 \; \mathrm{nm}$, giving $x \ge 7.9$.  Concerning photocatalytic Janus particles, work of Singh \textit{et al}. provides the main motivation for the present study.\cite{singh18} The incident UV light in that work has a wavelength $\lambda = 365\;\textrm{nm}$, and the ``small'' and ``large'' particles have diameters of $D=1\;\mathrm{\mu m}$ and $D=2\;\mathrm{\mu m}$, respectively, giving $x = 8.6$ and $x= 17.2$. These values are comparable to the size parameters in the previously mentioned studies. Furthermore, we note that an experimentalist who is specifically looking to reproduce the predictions of the present manuscript can use, for a given wavelength, particles of larger size, improving the validity of the geometrical optics approximation. }

Finally, we suggest two other promising directions of research. First, since the average velocity of a photo-active particle depends on light intensity, time-dependent illumination protocols could be used to program complex three-dimensional  trajectories. For this approach, one advantage is that the relative magnitudes of the light-parallel and light-perpendicular components of particle velocity are a function of light intensity; thus, the particle can be switched  between regimes of vertical motion and perpendicular motion through control over illumination (see Fig. \ref{fig:3dtraj}). Secondly, we suggest that our findings open a route towards light-tunable active fluid rheology. The rheological properties of a dilute suspension of active particles is governed by the probability distribution of particle orientations.\cite{saintillan18} For photo-active particles, we found that this distribution can be tuned by light intensity and direction.

\acknowledgments
W.E.U. acknowledges stimulating discussions with M. N. Popescu, L. G. Wilson, and D. P. Singh.  An early draft of this manuscript greatly benefited from comments from M. N. Popescu.  This research has benefited from scientific interactions facilitated by the COST Action MP1305 ``Flowing Matter'', supported by COST (European Cooperation in Science and Technology). W.E.U. thanks C. Pozrikidis for making the \texttt{BEMLIB} library freely available.\cite{pozrikidis02}
\\

\section{Appendix I. Coefficients $A_{lm}$}
\begin{flalign*}
A_{10}^{(lt90)} &= -\frac{1}{2 \pi} ((\pi - \alpha) \cos(\alpha) + \sin(\alpha)) & \\
A_{11}^{(lt90)} &= - \frac{1}{2 \pi} (\pi - \alpha) \sin(\alpha) &\\
A_{20}^{(lt90)} &=  \frac{5 }{32 } \cos^{2}(\alpha/2) (-1 + 3 \cos(\alpha)) &\\
A_{21}^{(lt90)} &= \frac{5 }{64} \csc^2 (\alpha/2) \sin^3(\alpha) &\\
A_{30}^{(lt90)} &=  \frac{7}{12 \pi} \sin^{3}(\alpha) &\\
A_{31}^{(lt90)} &= -\frac{7 }{24 \pi} \cos(\alpha) \sin^2(\alpha) &\\
A_{40}^{(lt90)} &= -  \frac{3 }{1024} \cos^{2}(\alpha/2) (-58 + 125 \cos(\alpha) &\\ 
& - 70 \cos(2 \alpha) + 35 \cos(3 \alpha))& \\
A_{41}^{(lt90)} &= -\frac{3 }{128} \cos^{3}(\alpha/2) \sin(\alpha/2) & \\ 
& (15 - 14 \cos(\alpha) + 7 \cos(2 \alpha)) &\\
A_{50}^{(lt90)} &=  - \frac{11 }{240 \pi}  (16 + 9 \cos(2 \alpha)) \sin^{3}(\alpha) &\\
A_{51}^{(lt90)} &= \frac{11 }{480 \pi} (7 \cos(\alpha) + 3 \cos(3 \alpha)) \sin^2(\alpha). & 
\end{flalign*}

\bibliography{shadowing}

\end{document}